\documentclass[11pt,a4paper]{article}
\usepackage{jheppub}

\usepackage{graphicx}
\usepackage{dcolumn}
\usepackage{bm}

\usepackage{booktabs}
\usepackage[english]{babel}
\usepackage{amsmath,amssymb,amsbsy,amstext, amsthm, simplewick}
\usepackage{graphicx}
\usepackage{amsfonts}
\usepackage{amssymb}
\usepackage{upgreek}
\usepackage{cancel}
\usepackage{color}
\usepackage{slashed}
\usepackage[dvipsnames]{xcolor}
\usepackage{tikz-cd}
\usepackage{feynmp-auto}
\usepackage{color}
\usepackage{soul}
\usepackage{dsfont}
\usepackage{verbatim}
\usepackage[utf8]{inputenc} 
\usepackage{soul}
\usepackage{epigraph}
\usepackage{todonotes}
\usepackage{csquotes}
\usepackage[normalem]{ulem}
\usepackage{subcaption}

\newcommand{\half}{\frac{1}{2}}

\def\beq{\begin{equation}}
\def\eeq{\end{equation}}
\def\bea{\begin{eqnarray}}
\def\eea{\end{eqnarray}}

\begin{document}

\title{
Global Structure, Non-Invertible PQ Symmetry, and the DFSZ Domain Wall Problem
}

\author[a]{Gongjun Choi,}
\author[b]{Sungwoo Hong,}
\author[c]{and Seth Koren}
\affiliation[a]{William I. Fine Theoretical Physics Institute,
University of Minnesota}
\affiliation[b]{Department of Physics, Korea Advanced Institute of Science and Technology}
\affiliation[c]{Department of Physics and Astronomy, University of Notre Dame}

\date{\today}

\abstract{In recent years it has become increasingly clear that the previously overlooked ``global structure'' of symmetry groups can encode significant theoretical structure and, more importantly, have substantial phenomenological implications. With this in mind we re-examine the DFSZ axion, which suffers from a domain wall problem due to the Standard Model generation structure.
We show that global structure $(G_{\rm EW} \times U(1)_{\rm PQ})/\mathbb{Z}_2$ acting between the Peccei–Quinn symmetry and the electroweak gauge group plays a crucial role in determining the precise nature of the domain wall problem, which has important implications in both cubic and quartic DFSZ.
We then demonstrate that the stability of the domain walls is enforced by a non-invertible chiral symmetry in quark flavor $Z'$ models which have additional global structure $(SU(3)_C \times G_F)/\mathbb{Z}_3$ acting between the color and the gauged quark flavor group.
The strategy of Non-invertible Naturalness then leads us to UV theories that resolve the domain wall problem through small-instanton-induced breaking of non-invertible symmetries. Finally, we sketch potential gravitational wave signatures arising from the annihilation of axion domain walls.
Our work illustrates the importance of considerations of global structure in realistic models of particle physics.
}

\maketitle

\makeatletter
\def\l@subsection#1#2{}
\def\l@subsubsection#1#2{}
\makeatother


\section{Introduction}
\label{sec:intro}

\subsection{Global Structure of Symmetry Groups in Particle Physics}

As particle physicists we usually think about small perturbations of quantum fields. After all, the particles of a theory are determined by such a perturbative analysis, which probes only the local structure of the field space. However, the physics of a theory is not fully characterized by perturbations around the vacuum, and the global structure of symmetry groups as well as the topology of field space can have important implications for phenomenology. Hence to fully understand theories of particle physics, it is necessary to determine how models with different global structures can be distinguished both theoretically and experimentally. Here `global structure' generally refers roughly to differences between symmetry groups which agree locally, but in this work we will consider only the differences between physics with symmetry group $G$ and with symmetry group a quotient $G/\Gamma$, where $\Gamma$ is a subgroup of the center of $G$. 

Importantly, this issue of global structure occurs in the Standard Model, where there are four possibilities for the gauge group
\begin{equation}
    G_{\rm SM} = \left(SU(3)_C \times SU(2)_L \times U(1)_Y\right)/\Gamma, \quad \Gamma = \lbrace 1, \mathbb{Z}_2, \mathbb{Z}_3, \mathbb{Z}_6\rbrace
\end{equation}
each consistent with all observed physics as they share the same Lie algebra \cite{Hucks:1990nw,Tong:2017oea}. This issue has garnered recent attention, in part due to the new technology and insights from the program of Generalized Global Symmetries \cite{Gaiotto:2014kfa}---which provides a gauge invariant characterization of the difference between these theories as they possess different global one-form symmetries. Connected to the different magnetic one-form symmetries, one physical difference is found in the spectrum of fractional instantons, which modifies the quantization conditions for axion couplings \cite{Reece:2023iqn,Choi:2023pdp,Cordova:2023her}. Connected to the different electric one-form symmetries, the representation theory of the groups differs in the possible existence of fractionally charged particles \cite{Alonso:2024pmq,Koren:2024xof,Koren:2025utp}. Intriguingly the topological responses of the theory can differ as well \cite{Hsin:2024lya,Wan:2024kaf}, though the phenomenology in this direction is less well-explored. Systematic group theory methods to determine all possible discrete ambiguities, including both the global structure and topological sector, in spontaneously broken gauge theory, such as the Standard Model and Grand Unified Theories, will be presented elsewhere \cite{Go:2026a, Go:2026b}. 

Aside from the Standard Model, there has been little investigation of issues of global structure in particle physics. Axion theories seem ripe for such issues to play important roles---as a compact scalar, the axion is the lower-form analog of a gauge field; its potential arises from instanton effects; its cosmology depends on the spectrum of cosmic strings and domain walls. In this work, we will consider the DFSZ model \cite{Dine:1981rt,Zhitnitsky:1980tq}, which is the canonical invisible axion model where the axion lives in fields which interact with the SM fermions. This model remains viable, but suffers from the domain wall problem (see next section) due to its close connection to the SM structure. In our analysis of this theory, the global structure of the symmetry groups and the topology of field space will have multiple important lessons for us. 
\begin{itemize}
    \item When identifying the axion mode as a direction in field space, it is necessary to remember that the space of global symmetries exists as an abstract manifold and the axion direction cannot depend on a choice of basis, which is often confused in the literature. (See Section \ref{sec:whosThatAxion})
    \item The cubic DFSZ model has overlooked $\mathbb{Z}_2$ global structure shared between the electroweak gauge group and the Peccei-Quinn symmetry, which impacts the spectrum of cosmic strings and domain walls. In particular this model has a $\mathbb{Z}_3$ domain wall problem rather than a $\mathbb{Z}_6$ domain wall problem, as has often been claimed. (See Section \ref{sec:global-gauge})
    \item If we gauge a subgroup of the approximate quark flavor symmetry of the SM which has shared global structure with $SU(3)_C$, then due to the existence of fractional instantons the $\mathbb{Z}_3$ subgroup of Peccei-Quinn symmetry is converted into a non-invertible symmetry. The strategy of non-invertible naturalness then leads us to an elegant solution to the domain wall problem. (See Section \ref{sec:gauge-gauge})
    \item The quartic DFSZ model legitimately has a $\mathbb{Z}_6$ domain wall problem, but if we embed this in a left-right model there is again $\mathbb{Z}_2$ global structure shared between the electroweak gauge group and the Peccei-Quinn symmetry. Now the domain wall problem is reduced to $\mathbb{Z}_3$, which can again become non-invertible and the domain wall problem can be fully resolved. (See Section \ref{sec:quartic})
\end{itemize}

That this simple, canonical BSM theory has such rich issues of global structure suggests that it is worth re-examining many ideas about BSM physics with these concepts in mind.

\subsection{The Domain Wall Problem}

The domain wall problem may occur whenever a discrete symmetry is spontaneously broken, and in particular plagues many well-motivated models of axions where the Peccei-Quinn symmetry is tied to the non-trivial structure of the Standard Model. We will get into the subtleties of this problem below, but in brief a Peccei-Quinn symmetry with a non-minimal anomaly coefficient $\mathcal{A} > 1$ yields an axion potential with $\mathcal{A}$ minima and one expects it to have $N_{\rm DW} = \mathcal{A}$ distinct domain walls. For $N_{\rm DW} > 1$ this results in the formation of a stable string-domain wall network. This occurs when the axion becomes massive and starts rolling down its potential which happens for $m\simeq H$, and different patches of spacetime can choose different, uncorrelated vacua. Domain wall network energy density, once it enters the scaling regime, is known to redshift slowly as $\rho_{\rm DW}\propto t^{-1}$, which corresponds to $\rho_{\rm DW}\propto a^{-2}$ in RD or $\rho_{\rm DW}\propto a^{-3/2}$ in MD. This in turn implies that any stable domain wall network eventually dominate the energy budget of the universe, in contrast with observed cosmology. 

One commonly discussed resolution is to assume a period of sufficiently low-scale inflation occurring after Peccei-Quinn symmetry breaking, such that any pre-existing topological defects are inflated away. This is indeed possible, but such a resolution relies on specific assumptions about the inflationary sector and its relation to the axion scale. It also poses further theoretical questions, as low scale inflation models are typically harder to achieve, requiring more fine-tuned flatness of the potential in order to fulfill slow-roll. In addition, the ratio $H_{\rm inf}/f_{a}$ needs to be suppressed enough to avoid overly large isocurvature perturbations~\cite{Linde:1985yf,Seckel:1985tj,Hertzberg:2008wr,Kawasaki:1995vt}. Considering these issues, it is highly desirable to explore alternative mechanisms that resolve the domain wall problem within the axion sector itself, ideally in ways that are theoretically well-motivated and accompanied by distinctive, experimentally testable consequences. Developing and systematically classifying such solutions is important not only for achieving a deeper and more complete understanding of axion models, but also for identifying new observational probes and discriminating signatures that can differentiate among axion scenarios.

One often-discussed possible solution to the domain wall problem is to add explicit breaking terms by hand to destabilize the domain wall network. In the presence of small explicit breaking effects, the axion acquires a ``bias potential'' in addition to the QCD potential, which lifts the vacuum degeneracy. In that case, while a domain wall network does form, pressure generated by the bias term eventually pushes the entire universe into the unique vacuum. While this approach is simple for a generic domain wall network, it requires extra care when it comes to models of the QCD axion for two reasons. Firstly, because the requirement that wall collapse not overproduce axion dark matter puts an upper bound on $f_a$ which is only an order of magnitude or so above the lower bound coming from neutron start cooling. Secondly, because a high-quality PQ symmetry in the UV phase is necessary to properly relax the strong CP angle. The issue is that an explicit PQ breaking term in the scalar potential generically comes with a complex parameter which induces a phase shift in the bias term compared to the QCD potential. Compatibility with the approximately CP-conserving vacuum imposes a very strong upper bound on the explicit breaking operator, while successful destruction of domain wall network at least before the onset of BBN, or before domain wall energy dominates the universe, puts a lower bound.

In \cite{Cordova:2023her} we recognized from a general, infrared analysis that a resolution to the domain wall problem could exist in axion models which contain non-invertible PQ symmetry.  It was further shown that the proposed mechanism based on non-invertible symmetry breaking does not occur in traditional grand unified theories, and no realistic model exhibiting this phenomenon has yet been written down. We will find that when flavorful gauge symmetries are considered, such a possibility does present itself.  Interestingly, we will see this connects naturally to realizations where the  additional explicit PQ breaking can be found to preserve the solution to the strong CP problem due to the automatic absence of a phase shift. A key aspect is the breaking being provided by ultraviolet instantons from a symmetry group that contains QCD as a subgroup.

There is another subtle and yet crucial point to make regarding the domain wall problem. Namely, the precise nature of the problem, and hence its solution, is not fully captured by the IR data and instead it is in general UV sensitive. To illustrate this point, consider an idealized situation where a set of axion experiments determine the axion potential to be of the form 
\beq
V(a) = \Lambda_{\rm QCD}^4 \left( 1 - \cos \left( N_{\rm DW} \frac{a}{f_a} \right) \right).
\eeq
IR physicists will conclude that there exists spontaneously broken $\mathbb{Z}_{N_{\rm DW}}$ shift symmetry, hence $\mathbb{Z}_{N_{\rm DW}}$ domain wall problem.\footnote{In fact there is an important earlier hurdle: just from measuring the axion mass and couplings $m_a, g_{a\gamma\gamma}, g_{aNN}$ one can only determine $N_{\rm DW}/f_a$, not either of them separately. Here we imagine this is somehow overcome, and our points is that still the domain wall problem is not understood from the IR.} However, it is still possible that, for example a subgroup $\mathbb{Z}_M \subset \mathbb{Z}_{N_{\rm DW}}$ can be effectively gauged and one such realization can occur when $U(1)_{\rm PQ}$ has non-trivial (discrete) global structure with gauge group.\footnote{Recently, gauging a discrete subgroup of axion shift symmetry via 4d BF theory was considered in \cite{Brennan:2023kpw} and resulting generalized symmetry structure and implications on topological defect physics were studied in detail.} The same applies when PQ symmetry has a discrete overlap with an exact global symmetry. As we analyze in the rest of the paper, in such case, the cosmic string--domain wall network structure can fundamentally differ from the naive $\mathbb{Z}_{N_{\rm DW}}$ picture. The key point is that the cosmological domain wall problem is really the question about the structure and dynamics of the string--wall network, and the actual version of the problem can really deviate from what the naive IR potential suggests: the precise form of the problem is UV sensitive. See \cite{Lu:2023ayc} for further examples and analysis of subtleties from this UV sensitivity.
It seems that unambiguous understanding of axion domain wall problem, which is central to the precise identification of underlying axion theory, requires either direct probe of UV degrees of freedom, or careful spectroscopy of axion defects.

\subsection{Non-invertible Naturalness}

In recent years, insights from generalized symmetries have shed new light on problems of particle physics e.g.~\cite{Cordova:2018cvg,Benini:2018reh,Cordova:2022qtz,Cordova:2022fhg,Cordova:2024ypu,Cordova:2022rer,Koren:2024xof,Koren:2025utp,Delgado:2026a,
Brennan:2020ehu,Cordova:2022ieu,Brennan:2022tyl,Cordova:2023ent,Cordova:2023her,Brennan:2023kpw,Brennan:2023tae, Brennan:2024iau,Brennan:2026ira,
vanBeest:2023dbu,vanBeest:2023mbs,
Davighi:2019rcd,Davighi:2020kok,Davighi:2024zip,Davighi:2024zjp,Davighi:2025awm,
Anber:2021upc,Anber:2024gis,Anber:2025gvb,
Cheung:2024ypq,
Cherman:2022eml,Cherman:2023xok,Chen:2024tsx,Chen:2025buv,
Hinterbichler:2022agn,Hinterbichler:2024cxn,
Hidaka:2020izy,Hidaka:2021mml,Hidaka:2021kkf,Yokokura:2022alv,Nakajima:2022feg,Hidaka:2024kfx,
Craig:2024dnl,
Choi:2022jqy,Choi:2022rfe,Choi:2022fgx,Choi:2023pdp,Choi:2025vxr,Burgess:2023ifd,Burgess:2025geh,Choi:2023gin,
Arbalestrier:2024oqg,
DelZotto:2024ngj,
Heidenreich:2020pkc,Heidenreich:2021xpr,Heidenreich:2023pbi, Fan:2021ntg,McNamara:2022lrw,Asadi:2022vys,Reece:2023iqn,Aloni:2024jpb,Reece:2024wrn,Garcia-Valdecasas:2024cqn,Reece:2025thc,
Wan:2019gqr,Wang:2020xyo,Wang:2021ayd,Wang:2021vki,Wang:2021hob,Wang:2022eag,Putrov:2023jqi,Wang:2023tbj,Choi:2025lkg,
Suzuki:2026xvf}. One insight from this technology has been the strategy of `non-invertible naturalness', which provides model-building guidance beyond 't Hooft's technical naturalness. A main shift in perspective is that a more-general notion of a system having a global symmetry is the existence of some surface operators having non-trivial but fully topological correlation functions---called `symmetry defect operators', these generalize familiar notions of Noether charges and Gaussian surfaces. This has clarified our understanding of symmetries of extended operators, which has exposed qualitative differences in how symmetries are explicitly broken between those acting on local or extended operators. This perspective also allows one to recognize non-group-like symmetries of systems which are difficult to recognize from the Lagrangian description alone. In particular, there are symmetries which act both on local operators and on extended operators, and these structures allow us to extract qualitatively new information about the breaking of such local operators.

The lesson is that by identifying a spurion for a non-invertible chiral symmetry, we learn that this parameter will be generated by nonperturbative gauge theory effects in an ultraviolet completion which includes the appropriate magnetically-charged states. While technical naturalness tells us that if some mechanism generated a small spurion in the UV, it will stay small into the IR, non-invertible naturalness goes further in guiding us toward a mechanism to generate the small spurion in the first place. We review the progress of model-building usage of non-invertible chiral symmetries briefly in Table \ref{tab:modelbuilding}.

\begin{table}[t]
    \centering
    \renewcommand{\arraystretch}{1.3}
    \begin{tabular}{|c|c|c|}
        \hline & Technically Natural & Unnatural\\\hline
        Dim. 0 & $y_\nu$ \cite{Cordova:2022fhg} & $\bar \theta$ \cite{Cordova:2024ypu} \\\hline
        Dim. 2 & $m_{12}^2$ \cite{Delgado:2024pcv} & The Hierarchy Problem\\\hline
        Dim. 4 & $\delta V(a)$ (here) & The CC Problem \\ \hline
    \end{tabular}
    \caption{The progression of model building utilizing the strategy of non-invertible naturalness, divided into the severity of the naturalness issue addressed and the dimension of the parameter protected by the non-invertible symmetry.}
    \label{tab:modelbuilding}
\end{table}

\begin{itemize}
    \item Neutrino masses can be spurions of non-invertible symmetries either with or without right-handed neutrinos, leading to theories of lepton flavor unification with instanton-suppressed Dirac or Majorana masses \cite{Cordova:2022fhg}.
    \item In the SM 1HDM the quark masses can be spurions of non-invertible symmetries, which leads to theories of quark color-flavor unification that can revive the simplest model of strong CP, the massless quark solution \cite{Georgi:128245,Kaplan:1986ru,Choi:1988sy} \cite{Cordova:2024ypu}.
    \item Adding a second Higgs doublet, the Higgs mass mixing $m_{12}^2$ can be a non-invertible spurion, leading to theories which revive the simplest axion model, the visible Peccei-Quinn Weinberg-Wilczek axion \cite{Peccei:1977hh,Peccei:1977ur,Weinberg:1977ma,Wilczek:1977pj} \cite{Delgado:2024pcv}.
\end{itemize}

The current analysis of the DFSZ model \cite{Dine:1981rt,Zhitnitsky:1980tq} is the natural continuation of these previous works, where we now proceed to the invisible axion and add both the second Higgs doublet and a complex scalar singlet. Two different versions of the DFSZ model exist depending on whether the singlet scalar is coupled to the two Higgs doublets with a cubic or a quartic interaction. In both cases, we will show that the global structure between PQ and the electroweak gauge group plays a central role in the precise identification of domain wall problem. In Section \ref{sec:cubic} we will find the domain wall problem of the cubic model amenable to a non-invertible symmetry-based solution, and in Section \ref{sec:quartic} we will elegantly extend the quartic model to include the same structure. In Section \ref{sec:CFunification} we will embed both of these theories in ultraviolet theories of quark color-flavor unification, providing the non-invertible symmetry-breaking to lift the vacuum degeneracy. In Section \ref{sec:GW} we will briefly discuss the gravitational wave cosmology of these models.

In general, theories of axions have been shown to have particularly rich generalized symmetry structures, including higher-form symmetries \cite{Banks:2010zn,Gaiotto:2014kfa}, lower-form symmetries \cite{Aloni:2024jpb,Anber:2024gis}, non-invertible zero-form symmetries \cite{Cordova:2022ieu,Cordova:2023her,DelZotto:2024ngj,Choi:2022fgx,Delgado:2024pcv,Choi:2025vxr,Hong:2025qbw}, higher group symmetries \cite{Brennan:2020ehu,Hidaka:2020izy,Hidaka:2021mml,Hidaka:2021kkf,Nakajima:2022feg,Brennan:2023kpw,Anber:2024gis}, and non-invertible higher-form symmetries \cite{Yokokura:2022alv,Choi:2022fgx,Hidaka:2024kfx,DelZotto:2024ngj,Hong:2025qbw}.

\section{Cubic DFSZ} \label{sec:cubic}

\begin{table}\centering
\large
\renewcommand{\arraystretch}{1.3}
\begin{tabular}{|c|c|c|c|c|c|}  \hline
 & $SU(3)_C$ & $SU(2)_L$ & $U(1)_Y$ & $U(1)_{Q-N_cL}$ & $U(1)_{\rm PQ}$ \\ \hline

$Q$ & $3$ & $2$ & $+1$ & $+1$ & $+1$  \\ \hline

$\bar u$ & $\bar 3$ & $-$ & $-4$ & $-1$ & $0$ \\ \hline

$\bar d$ & $\bar 3$ & $-$ & $+2$ & $-1$ & $0$\\ \hline

$H_u$ & $-$ & $2$ & $+3$ & $0$ & $-1$ \\ \hline

$H_d$ & $-$ & $2$ & $-3$ & $0$ & $-1$ \\ \hline

$\phi$ & $-$ & $-$ & $0$ & $0$ & $+2$ \\ \hline

\end{tabular}\caption{Symmetries in the low-energy cubic DFSZ model.}\label{tab:irchargesCubic}
\end{table}

The DFSZ model upgrades the scalar sector of the SM with the standard type II two Higgs doublets $H_u, H_d$ and a complex scalar $\phi$, and writes the following interaction terms which constrain the zero-form global symmetries:
\begin{equation} \label{eqn:yukawas}
    \mathcal{L} \supset (y_t)^i_j H_u Q_i \bar u^j + (y_b)^i_j H_d Q_i \bar d^j + \lambda_3 H_u H_d \phi + {\text{ h.c.}} 
\end{equation}
The leptons can be included straightforwardly, but we leave them out henceforth as they will not play an important role in the physics we will discuss. This theory has multiple $U(1)$ gauge and global symmetries, and carefully considering this space, while properly taking into account the global structure, will be the subject of much of this work. Let us first give a brief discussion on symmetries of the theory.

We start by counting the number of $U(1)$ symmetries.\footnote{Throughout this work, we take care to work with $U(1)$ symmetries whose least charged excitation has charge 1. Ultimately this is merely a useful convention, as the physics must be independent of the normalization of the charges---there is nothing intrinsically wrong with a $U(1)$ symmetry with non-integer charges so long as they are relatively rational, as in the usual normalization of hypercharge in the Standard Model. However, when departing from this convention various topological aspects of the physics are easily obscured and in practice it is easy to make mistakes. This convention ensures, for example, that symmetry rotations by $2\pi$ act as the identity; that anomaly coefficients are integers; and that Dirac indices of fermions correspond to the integer number of zero modes in an instanton background. \label{foot:u1charge}}
First, any generation-dependent symmetries are explicitly broken by the Yukawa couplings $y_t$ and $y_b$.
Then the gauge theory of the fields appearing in \eqref{eqn:yukawas} has $U(1)^6$ zero-form symmetry. Three perturbative interactions in \eqref{eqn:yukawas} reduce this to three $U(1)$s, one of which is gauged as $U(1)_Y$. Therefore, we expect two global $U(1)$ symmetries, possibly with ABJ anomalies. 

The gauge quantum numbers of the fields are uniquely fixed by the couplings to the gauge boson through the covariant derivative, and we have the gauged $U(1)_Y$ hypercharge such that $H_u, H_d$ have opposite charges and $\phi$ is a singlet.
It is important to realize that global symmetry charges are not uniquely determined, but entail a choice of basis, one choice of which we give in Table \ref{tab:irchargesCubic}.   
As is familiar from the Standard Model, there is a generation-independent, ABJ anomaly-free Abelian symmetry under which the fermions transform nontrivially, which it is often useful to discuss as `quark number minus $N_c$ times the lepton number' $U(1)_{Q-N_cL}$ and after confinement acts faithfully as `baryon minus lepton number' $B-L = (Q-N_c L)/N_c$. In this choice of basis the scalars are all uncharged, but note that we could equally as well take some linear combination with hypercharge $Q' = k_1 (Q - N_cL) + k_2 Y$, $k_i \in \mathbb{Z}$, and this would remain a generation-independent, ABJ anomaly-free Abelian symmetry.

Just as the Yukawas explicitly break many of the SM flavor symmetries, the coupling $\lambda_3$ explicitly breaks one would-be $U(1)$ direction. In particular for $\lambda_3 = 0$ we decouple $\phi$ from the SM fields---in such a theory $\phi$ has its own $U(1)$ global symmetry, and the remaining 2HDM consists of the Peccei-Quinn Weinberg-Wilczek model of the visible axion, so has a $U(1)_{\rm PQ}$ symmetry. With finite $\lambda_3$ we have the DFSZ model of the invisible axion, and one of these directions is explicitly broken, leaving a $U(1)_{\rm PQ}$ symmetry under which $\phi$ is charged. As we will discuss further below, the Peccei-Quinn symmetry is not uniquely fixed, as any addition of anomaly-free global or gauged $U(1)$ charges does not modify the anomaly. The zero-form symmetries are abstractly dictated by the Lagrangian of \eqref{eqn:yukawas} (in full, the partition function of the theory with this Lagrangian) and so the physics is necessarily invariant under such shifts in what we call each symmetry.

\subsection{Who's That Axion?} \label{sec:whosThatAxion}

The identification of the axion mode in models with multiple $U(1)$ gauge or global symmetries has been often confused throughout the literature. Indeed there are a wide variety of subtleties, and this topic is ripe for further careful examination. 

An interesting aspect of axion physics is the interconnected roles of the scalar sector and the fermion sector. As the axion mode is contained within the pseudoscalar phases of the theory, the identity of the axion is concerned mainly with the physics of the scalar sector. On the other hand, the quantized couplings of the axion to the instanton densities come from the fermion sector, while the two sectors are linked by the yukawa terms (in a renormalizable theory).

We seek to identify the axion as a particular linear combination of the pseudoscalar phases of complex scalar fields with vevs. 
What should we get out from such an identification? Let us list various aspects of the axion identity, which we do not claim form a minimal necessary list, or even a sufficient list in a more general case.

\begin{itemize}
    \item The axion $a$ is a dimension-one scalar field that transforms nonlinearly under the $U(1)_{\rm PQ}$ Peccei-Quinn symmetry, $a \rightarrow a + \alpha_{\rm PQ} f_a$, where $\alpha_{\rm PQ}$ is the rotation angle and $f_a$ is known as the decay constant. 
    \item As implied by the first statement, in the limit that $U(1)_{\rm PQ}$ is exact, the axion potential is flat---it is the massless Goldstone boson from the spontaneous breaking of this symmetry by one or more complex scalar fields with nonzero vevs. 
    \item A `Peccei-Quinn symmetry' has an ABJ anomaly with $SU(3)_C$, which leads to a Chern-Simons-type coupling to the gauge field strength $F_C$, of  $\mathcal{L} \supset \frac{\mathcal{A}}{f_a} \frac{1}{8\pi^2 } \text{tr}(a \wedge F_C \wedge F_C)$, where $\mathcal{A}$ is the PQ-$SU(3)_C^2$ anomaly coefficient. 
    \item This implies that the axion may not transform under continuous gauge symmmetries (such as hypercharge), which would ruin the gauge invariance of this term. However, it is allowable for the axion to transform under discrete gauge symmetries, as long as $a \rightarrow a + 2\pi/k$ where $k|\mathcal{A}$.
    \item `The' Peccei-Quinn symmetry is defined only up to the addition of generators of other non-anomalous $U(1)$ gauge and global symmetries, which do not change any of the properties above. Then the description of $a$ in terms of scalar phases \textit{must} be invariant under such redundancies of description, which provides an important consistency check on any putative identification of such a mode.
    \item As the goldstone of a compact symmetry, the axion is a compact field and the decay constant $f_a$ can be interpreted as the size of the field space, in that $\oint_\gamma \frac{da}{2\pi} = k f_a, \ k \in \mathbb{Z}$ for any curve $\gamma$ ($k = 0$ for any contractible curve). The size of the field space is not directly physical, though the Chern-Simons coupling is: $f_a$ is defined only up to reparametrizations of the Peccei-Quinn symmetry that change the anomaly coefficient $\mathcal{A}$ (for example, rescaling all of the PQ charges uniformly). 
    \item The axion shift symmetry is explicitly broken by QCD instanton effects due to the Chern-Simons coupling, as well as by the general nonperturbative effects of confinement, which constitute the largest contribution to the axion potential. This potential gives the axion a mass and breaks the continuous shift symmetry---however these effects may preserve a discrete shift symmetry, which can lead to the domain wall problem as we will discuss below.

\end{itemize}

The identification of the axion mode is simple enough in the KSVZ model (with one relevant scalar) or the PQWW model (with two relevant scalars), but for the DFSZ model this becomes somewhat nontrivial. 
Here the axion mode is a combination of the phase modes of $H_u, H_d, \phi$, which all condense with vacuum expectation values (vevs) $\langle H_u \rangle = \frac{v_u}{\sqrt{2}} \equiv \frac{v_{\rm EW}}{\sqrt{2}} \sin \beta, \langle H_d\rangle \equiv \frac{v_d}{\sqrt{2}} = \frac{v_{\rm EW}}{\sqrt{2}} \cos\beta, \langle \phi \rangle \equiv \frac{v_\phi}{\sqrt{2}}$, where $v_{\rm EW}^2 = v_u^2 + v_d^2$ and $\tan\beta=v_{u}/v_{d}$ as familiar in 2HDMs. Each of these has one electrically neutral pseudoscalar mode, and among them they display the varied outcomes one may have for such a compact direction: One direction will be gauged and so exact, and thus one combination will become the longitudinal mode of the $Z$ boson; one direction will be explicitly broken by the cubic interaction, so another combination will be the massive pseudoscalar mode of the 2HDM; one direction will be classically good but suffers an ABJ anomaly with QCD, so a final combination will be the axion.

We expand each field around its vev in a nonlinear representation for the neutral pseudoscalar modes, ignoring the others to simplify the discussion
\begin{equation} \label{eq:cubicPhases}
    H_u = \begin{pmatrix}
        0 \\ \frac{v_u}{\sqrt{2}} \\
    \end{pmatrix} e^{i \theta_u}, \qquad H_d = \begin{pmatrix}
        \frac{v_d}{\sqrt{2}}  \\ 0 
    \end{pmatrix} e^{i \theta_d}, \qquad \phi = \frac{v_\phi}{\sqrt{2}} e^{i\theta_\phi}.
\end{equation}
The neutral pseudoscalar field space is the 3-torus $T^3$, and we seek an invariant identification of the hypercharge longitudinal mode and the axion, which are guaranteed by Goldstone's theorem to be compact directions.
The direction which is given a large mass by the explicit breaking of the third would-be $U(1)$ in the scalar sector is simply proportional to $\theta_u + \theta_d + \theta_\phi$, the combination of phases which enter the $\lambda_3$ term, and we will return to this at the end. 

The canonically normalized pseudoscalar modes we have started with can be arranged into the dimension one vector 
\begin{equation}
    \vec{\vartheta} = (v_u \theta_u, v_d \theta_d, v_\phi \theta_\phi),
\end{equation}
which combines the neutral phases with the vevs of the radial modes. The modes we wish to construct will then have the form 
\begin{equation}
    a = \vec{\vartheta} \cdot \hat{a}, \qquad y = \vec{\vartheta} \cdot \hat{y}
\end{equation}
where $a$ is the axion mode, $y$ is the longitudinal mode of the $Z$ boson, and $\hat{a}, \hat{y}$ are dimensionless unit vectors in three dimensional field space and result in canonically normalized $a, y$ when $\hat{a} \cdot \hat{a} = 1 = \hat{y} \cdot \hat{y}$.\footnote{We note that our constructions in this section make use of a baby version of field space geometry where we have assumed the canonical form of the kinetic terms of $H_u, H_d, \phi$ and so contractions are performed with the Euclidean metric. A more-general determination of the axion mode would benefit from being framed explicitly in terms of the field space metric. See \cite{Buen-Abad:2021fwq} for some perspective in this direction.} We should construct these out of variables which transform covariantly under redefinitions of the global symmetries, in particular we have the dimension one charge vectors
\begin{equation}
    \vec{X} = (x_u v_u, x_d v_d, x_\phi v_\phi), \qquad \vec{Y} = (y_u v_u, y_d v_d, y_\phi v_\phi),
\end{equation}
where $x_i$ are PQ charges and $y_i$ are hypercharges. It will be useful also to define the unit vectors $\hat{Y} = \vec{Y}/|\vec{Y}|,\  \hat{X} = \vec{X}/|\vec{X}|$.
Under a hypercharge rotation by $\alpha_Y$ and a PQ rotation by $\alpha_{\rm PQ}$, the vector of phases rotates as
\begin{equation}
    \vec{\vartheta} \rightarrow \vec{\vartheta} + \alpha_Y \vec{Y} + \alpha_{\rm PQ} \vec{X},
\end{equation}
while under a change of basis for the global symmetry, the PQ charge vector undergoes a change
\begin{equation}
    \vec{X} \rightarrow \vec{X} + c \vec{Y}, \quad c \in \mathbb{Z}.
\end{equation}
This covariant language clarifies which combinations of phases are sensible to consider, which allows the correct modes to be determined just from considerations of symmetry transformations and reparametrization invariance.

The gauge symmetry, unlike the global symmetry, is uniquely determined by the coupling of the fields to the gauge boson, so that the longitudinal mode is simply $\hat{y} = \hat{Y}$,
\begin{equation} \label{eq:cubicHypercharge}
    y \equiv  \vec{\vartheta} \cdot \hat{Y} = \frac{1}{v_{\rm EW}} (v_u^2 \theta_u - v_d^2 \theta_d).
\end{equation}
It is straightforward to check that under a hypercharge transformation by $\alpha_Y= 2\pi$, $y$ shifts as $y \mapsto y + 2\pi f_Y$, as desired, where the size of the field space is $f_Y^2 \equiv | \vec{Y}|^2 = 9 v_{\rm EW}^2$.\footnote{We can see this directly by looking at the kinetic terms of the charged scalars, which lead to $\mathcal{L} \supset |d \vec{\vartheta} - \vec{Y} A|^2$, such that a gauge transformation by $\lambda =  -\vec{\vartheta}\cdot \hat{Y}/f_Y = - y/f_Y$ implements unitary gauge.}

The naive axion mode would be similarly $\vec{\vartheta} \cdot \hat{X}$, but this mode would inappropriately transform nonlinearly under hypercharge since $\vec{X}$ is not orthogonal to $\vec{Y}$. We can construct an appropriately orthogonal linear combination by simply subtracting the component of $\vec{X}$ along $\vec{Y}$ and then re-normalizing, which gives\footnote{This issue was identified in a classic paper of Srednicki \cite{Srednicki:1985xd}, though the solution was taken to imply the necessity of irrational PQ charges. Our covariant approach ensures invariance under redundancies in description and hopefully clarifies how to approach axion identification more generally.}
\begin{equation} 
    \hat{a} = \frac{1}{f_a}\left( \vec{X} - (\vec{X}\cdot \hat{Y}) \hat{Y}\right), \qquad f_a^2 = \left| \vec{X} - (\vec{X}\cdot \hat{Y}) \hat{Y}\right|^2 \nonumber  
\end{equation}
\begin{equation}\label{eqn:cubicAxMode}
a \equiv \vec{\vartheta} \cdot  \hat{a}= \frac{2}{f_a}\left[v^2_\phi \theta_\phi - \frac{v_u^2 v_d^2}{v_{\rm EW}^2} \left(\theta_u + \theta_d \right) \right].
\end{equation}

Perhaps the most crucial aspect of this covariant construction is that it makes manifest the invariance under different basis choices for the global symmetry $\vec{X} \rightarrow \vec{X} + c \vec{Y}$. The orthogonality with hypercharge $\vec{Y}\cdot\hat{a}=0$ ensures that under Peccei-Quinn and hypercharge symmetry transformations we have only $a \rightarrow a + \alpha_{\rm PQ} f_a$, so the axion is not charged under hypercharge.\footnote{Note that since we have already specialized above to consider only the $U(1)_{\rm EM}$-neutral pseudoscalar phases, their $U(1)_Y$ and $T_3^L \subset SU(2)_L$ transformations are linked, such that the axion is now invariant under both. One would ultimately like to start from the fully abstract scalar field space and show how the dynamics invariantly pick out the axion direction. This could be necessary to fully account for topological effects in a general theory, but we leave this important task for future work.} For the $y$ mode, $ y \rightarrow y + \alpha_Y f_Y + \alpha_{\rm PQ} \frac{v_d^2 - v_u^2}{v_{\rm EW}}$. This will imply that some of the axion strings contain fractional $Z$-boson magnetic flux, and we refer to \cite{Abe:2020ure} for detailed construction of some DFSZ cosmic strings (in the quartic model) and \cite{Niu:2023khv} for related general issues.

Finally we return to the explicitly broken direction $A$, which gains a mass proportional to $\lambda_3$. This coupling leads to a pseudoscalar potential
\begin{equation}
    V\supset \lambda_3 H_u H_d \phi + \text{h.c.} \supset  \frac{1}{\sqrt{2}} \lambda_3 v_u v_d v_\phi \cos\left( \theta_u + \theta_d + \theta_\phi\right),
\end{equation}
and we can read off the massive direction. In terms of canonically normalized scalar modes, this is 
\begin{equation}
    A = \vec{\vartheta} \cdot \hat{A}, \qquad \hat{A} = c_A \left(\frac{1}{v_u}, \frac{1}{v_d}, \frac{1}{v_\phi}\right), \qquad c_A^2 = \frac{v_u^2 v_d^2 v_\phi^2}{v_d^2 v_\phi^2 + v_u^2 v_\phi^2 + v_u^2 v_d^2}.
\end{equation}
This mass eigenvector direction points in a generic angle on the field space torus which winds around densely, although of course the field space is compact. Resultingly, $A$ is not a periodic scalar, implying that its coupling to the instanton density may not be integer quantized.

This explicitly broken mode is orthogonal from the others by construction, as 
\begin{equation}
    \hat{A} \cdot \vec{Y} \propto y_u + y_d + y_\phi=0, \qquad \hat{A} \cdot \vec{X} \propto x_u + x_d + x_\phi=0.
\end{equation}
That these particular sums vanish is exactly the constraint placed on the space of good classical symmetries by the existence of the $\lambda_3$ term in the Lagrangian, i.e.~$\lambda_3$ operator is both hypercharge and PQ neutral.

Thus we have succeeded in constructing a useful basis of the canonically normalized pseudoscalar modes $\left(a, y, A\right)$ where $A$ has a classical mass, $y$ is eaten, and $a$ is the axion mode.\footnote{In the literature one finds errors which tend to stem from confusion over the notion of `orthogonality' to impose. For example, an often-invoked condition on \textit{charges}, such as $(x_u, x_d, x_\phi) \cdot (y_u, y_d, y_\phi) = 0$, has no invariant meaning whenever one might have the freedom to impose such a condition, instead obscuring the symmetry reparametrization invariance. Also, contrary to many claims, it poses no issue that the longitudinal mode of the $Z$ boson manifestly transforms under PQ which merely implies the existence of magnetic fluxes through axion strings---indeed there is no freedom to prevent this.  Furthermore, Eq.~\ref{eqn:cubicAxMode} is sometimes understood as a modification of the PQ charges of the theory to $\vec{X}' = \vec{X} - \vec{X} \cdot \hat{Y} \hat{Y}$. Practically this does not make sense because it leads generically to relatively irrational charges for the fields, which would have the effect of `decompactifying' $U(1)_{\rm PQ}$ to $\mathbb{R}_{\rm PQ}$. While the local physics is not sensitive to this, the breaking of $\mathbb{R}_{\rm PQ}$ would not yield cosmic strings. It is not apparent to us the extent to which this has lead to inaccuracies in the literature---if one solves for cosmic strings at the level of all the scalar fields and imposes that the configurations are smooth, then the physics of the strings should remain correct. However, the connection to the axion mode and the PQ symmetry is obscured.} This tells us how to decompose any pseudoscalar mode into a linear combination of these basis vectors, as will be useful in the next section when considering the anomalous direction which appears coupled to the QCD instanton $\propto \text{tr}(F_C \wedge F_C)$.

\subsection{ABJ Anomaly and Naive Domain Wall Number}

The ABJ anomaly coefficient for the PQ symmetry defined in Table \ref{tab:irchargesCubic} in an $SU(3)_C$ instanton background can be easily computed as
\begin{equation}
    \mathcal{A}_{3} = 2 N_g,
\end{equation}
leading to an anomalous conservation equation for the $U(1)_{\rm PQ}$ current 
\begin{align}
    \partial_\mu J^\mu_{\rm PQ} &= \frac{\mathcal{A}_{3}}{16\pi^2} \text{Tr }F_C \tilde{F}_C \\
    \Delta Q_{\rm PQ} &= \mathcal{A}_{3} \mathcal{N}_C
\end{align}
where $F_C$ is the color field strength, $\tilde{F}_C$ is its Hodge dual, and $\mathcal{N}_C = \frac{1}{16\pi^2} \int \text{Tr }F_C \tilde{F}_C$ is the instanton number for a given $SU(3)_C$ configuration. 

As to the coupling of the axion to the gauge fields, we may follow the same procedure as is normally done in KSVZ-type models to remove the pseudoscalar phases from the fermion mass terms using anomalous field redefinitions.\footnote{Parenthetically, we note that discussions of this procedure in the literature often refer to these field redefinitions as Peccei-Quinn transformations, which we find confusing. They are explicitly not $U(1)_{\rm PQ}$ transformations, under which these terms are invariant because  $U(1)_{\rm PQ}$ is a good classical symmetry. Instead these anomalous field redefinitions of the PQ-charged fermions move the PQ dependence of the theory from the fermion sector into the topological density term.} We take the Lagrangian after spontaneous symmetry breaking,
\begin{equation}
    \mathcal{L} \supset (m_u)^i_j \ e^{i \theta_u} \ u_i \bar u^j + (m_d)^i_j \ e^{i \theta_d} \ d_i \bar d^j, 
\end{equation} 
Then we perform anomalous field redefinitions $u_i \mapsto u'_i = e^{i \theta_u} u_i$, $d_i \mapsto d'_i = e^{i \theta_d} d_i$, which remove the phases from the mass matrices and render $u_i', d_i'$ invariant under PQ transformations. Due to the change in the path integral measure under each of these, an extra term is generated in the Lagrangian
\begin{equation}
    \mathcal{L} \supset N_g (\theta_u + \theta_d) \frac{\text{Tr }F_C \tilde{F}_C}{16\pi^2},
    \label{eq:QCD_CS_term}
\end{equation}
in addition to derivative couplings of the phases to the fermions. We can use the results of the previous section to translate this into the physical basis, finding 
\begin{align}
    \theta_u + \theta_d &= \vec{\vartheta} \cdot \left(\frac{1}{v_u},\frac{1}{v_d},0\right) \\
    &= \vec{\vartheta} \cdot \left[- \frac{2}{f_a} \hat{a} + 0 \hat{y} + \frac{4 v_\phi^2}{c_A f_a^2}\hat{A}\right].
\end{align}
We see the coupling of the axion to the gauge fields is indeed quantized 
\begin{equation}
    \mathcal{L} \supset 2 N_g \frac{a}{f_a} \frac{\text{Tr }F_C \tilde{F}_C}{16\pi^2}, 
\end{equation}
which shows the ABJ anomaly explicitly breaks $U(1)_{\rm PQ} \rightarrow \mathbb{Z}_{2 N_g}$.\footnote{One may be worried by the irrational coefficient in front of $A$ term. Recall, however, that $A$ as a dimension one field embedded in $T^3$ field space does not correspond to a periodic field. Regardless, Eq.~\eqref{eq:QCD_CS_term} is perfectly consistent with $2\pi$ periodicity of $\theta$s, although it is obscured once it is decomposed in terms of the mass eigenstates.}
 
Flowing below the QCD scale, and having integrated out the massive $A$, this coupling is responsible for generating a potential for the axion as a result of the explicit breaking of $U(1)_{\rm PQ}$ due to the anomaly
\begin{equation} \label{eq:cubicAxPot}
    V(a) = \Lambda_{\rm QCD}^4 \left[1 - \cos\left(2 N_g \frac{a}{f_a}\right)\right]. 
\end{equation}
In the early universe this potential is generated dynamically at temperatures $T \sim \Lambda_{\rm QCD}$. The axion mass $m_a \sim \Lambda_{\rm QCD}^2/f_a$  is already comparable to the Hubble expansion rate $H \sim T^2 / M_{\rm pl}$, so the axion thereafter settles into different, disconnected vacua in different regions of the universe. This constitutes the formation of a domain wall network separating regions of different vacua.

Naively, the anomaly coefficient $\mathcal{A}_{3}$ would suggest the presence of $2N_g$ distinct, degenerate vacua labeled by $\langle a/f_a \rangle = 2\pi k/\mathcal{A}_{3}$, $k = 0, \cdots, \mathcal{A}_{3} - 1$. Then the induced axion potential $V(a)$ results in spontaneous breaking of 0-form $\mathbb{Z}_{2Ng}$ axion shift symmetry, implying the formation of $N_{\rm DW}=2N_g$ domain walls. Resultingly, the anomaly coefficient $\mathcal{A}_{3}$ is often referred to as the `domain wall number' $N_{\rm DW} = \mathcal{A}_{3}$ in axion literature. If we further consider an axion string around which the axion winds from $a/f_a = 0$ to $2\pi$ this will then be attached to $N_{\rm DW}$ domain walls. However, as we show below, this conclusion is imprecise, and a correct analysis must take into account the ``global structure'' between the global symmetry group and gauged symmetry group: in the current theory, there exists $\mathbb{Z}_2$ overlap between $U(1)_{\rm PQ}$ and $SU(2)_L$ and/or $U(1)_Y$. Taking into account the non-trivial global structure $(G_{\rm SM} \times U(1)_{\rm PQ}) / \mathbb{Z}_2$, the cosmic string - domain wall structure fundamentally changes. We show below that in the DFSZ model this results in the reduction of $N_{\rm DW}$ to $N_{\rm DW} = \mathcal{A}_{3} / 2$.

When $N_{\rm DW}$ is equal to 1, a single domain wall ends on a string, whose tension causes it to contract and disappear quickly. In contrast, when there are multiple domain walls ending on a string, a string-wall network is formed, which can be stable. 
Once the system enters the scaling regime, the energy density of the domain wall network gets diluted as $\rho_{\rm \scriptscriptstyle DW}\propto t^{-1}$, which is slower than radiation and matter. Thus, the stable domain wall will eventually overclose the universe, causing anisotropic expansion of the universe and inconsistency with observations from CMB and large-scale structure~\cite{Zeldovich:1974uw,Vilenkin:1984ib,Sikivie:1982qv} (see \cite{Hong:2025piv}, however, for the possibility of domain wall dominance in the early universe and associated gravitational wave signals). This is the domain wall problem, and the analysis only based on the ABJ anomaly data would suggest that $N_{\rm DW}=2 N_g$ and the cubic DFSZ model naively suffers a $\mathbb{Z}_6$ domain wall problem. As reviewed in Section~\ref{sec:intro}, the axion domain wall problem constitutes a genuine and potentially fatal challenge for axion theories and must therefore be taken seriously: any viable axion model should be free of stable domain walls that would otherwise dominate the energy density of the Universe and contradict cosmological observations. 

In the rest of this paper, we discuss the resolution to the domain wall problem in two distinct steps. Firstly, as alluded above already, it is crucial that the analysis of string-domain wall network must incorporate potential global structure of symmetry groups. In particular, identification of the fully correct structure of string-domain wall network is possible by taking into account $\mathbb{Z}_2$ global structure between \emph{global} $U(1)_{\rm PQ}$ and \emph{gauged} $SU(2)_L \times U(1)_Y$ (potentially with yet another $\mathbb{Z}_2$ quotient between the two electroweak gauge group factors, but inclusion of this does not change our conclusion). This first step allows us to establish the precise definition of the DFSZ domain wall problem. This also means that the cubic DFSZ model in fact implements a $\mathbb{Z}_2$ Lazarides-Shafi mechanism \cite{Lazarides:1982tw}, as we will discuss further below, which reduces the domain wall problem from the naive $2N_g$ to $N_g$. We emphasize that we did not need to add or modify anything to realize this. The original DFSZ theory already has this structure present. We only needed to understand underlying, so far overlooked, global structure to reveal the correct physics. 

Our second step to completely solve the DFSZ domain wall problem is to introduce a second type of global structure. Specifically, we extend the SM by gauging the approximate $SU(3)_H$ flavor symmetry of the quark sector. Crucially, the feature $N_C = N_g$ of SM makes it possible to introduce a $\mathbb{Z}_3$ quotient of the form $(SU(3)_C \times SU(3)_H)/ \mathbb{Z}_3$, which is global structure between two gauge symmetries. Then we will realize that in this simple gauged flavor extension of the SM, the $\mathbb{Z}_3$ subgroup of $U(1)_{\rm PQ}$ becomes a \emph{non-invertible} chiral symmetry. In \cite{Choi:2023pdp, Cordova:2023her} it was shown that global structure of gauge group can result in non-invertible axion shift symmetry,\footnote{See \cite{Agrawal:2022lsp,Agrawal:2024ejr,Agrawal:2025rbr,Benabou:2025kgx,Reig:2025dqb} also for axion-gauge couplings in Grand Unified Theories and string theory. Also, see \cite{Hong:2025qbw} for more details on non-invertible chiral symmetries arising from $U(1)$, fractional, or general CFU-type instantons in 4d QFT.} and \cite{Cordova:2023her} further discussed that ``non-invertible axion domain wall problem'' can be solved by implementing non-invertible symmetry breaking obtained by a UV completion that generates small instantons. Motivated by this, we will thus solve the domain wall problem of cubic DFSZ by embedding that model in an ultraviolet theory that has appropriate dynamical monopoles (or small instantons) to implement non-invertible symmetry breaking.

\subsection{Global-Gauge Global Structure and Cosmic Strings} \label{sec:global-gauge}

\paragraph{Warm up: charge rescaling} The prescription for identifying $N_{\rm DW}$ as simply $N_{\rm DW} = \mathcal{A}$ is obviously incomplete. Above we accounted for changes of basis which rotate other anomaly-free symmetry directions into PQ, and $\mathcal{A}$ does not change under such a redefinition. But we could also consider redefinitions which do change the anomaly, for example a PQ' which simply multiplies all the PQ charges by non-zero $k \in \mathbb{Z}$. 

Under such a basis change the axion direction appropriately does not change, but the anomaly coefficient scales linearly as $\mathcal{A} \rightarrow k \mathcal{A}$. This is compensated by a change in the size of the axion field space, $f_a \rightarrow |k| f_a$, such that the overall Chern-Simons coupling of the axion to the instanton density is invariant $\mathcal{L} \supset \mathcal{A} \frac{a}{f_a} {\rm tr} \frac{F_C \wedge F_C} {8\pi^2}$.

So the size of the axion field space is not itself defined absolutely. However, it is natural to define the PQ symmetry such that the particle with smallest charge has charge $1$ (though it's possible the relevant excitation is a composite, for example of charge $3$ and charge $-2$ particles). This ensures that the full $U(1)$ symmetry acts faithfully on the spectrum of the theory---if we choose to instead multiply by $k$, we end up with a $U(1)/\mathbb{Z}_k$ faithfully acting symmetry. In such a basis, the $\mathbb{Z}_k$ subgroup of the PQ symmetry cannot be broken (since nothing transforms under this subgroup), so this factor manifestly does not contribute to the domain wall problem. 

Let us briefly work through this example in the simple case with a single scalar field and a single KSVZ species: $\mathcal{L} \supset \psi \phi \bar\psi$, where $\langle \phi\rangle = \frac{v_\phi}{\sqrt{2}} e^{i \theta_\phi}$. There is a PQ symmetry which rotates $\phi$ and $\psi$ oppositely, and we can assign them an arbitrary charge $k$. The technology of Section \ref{sec:whosThatAxion} is simple in this case and tells us $f_a = k v_\phi$, $a = \theta_\phi f_a/k$ so that under a PQ transformation by $\alpha$, we have $\theta_\phi \rightarrow \theta_\phi + k \alpha$ and so $a \rightarrow a + f_a \alpha$.

When we rotate the axion out of the mass matrix, we get a coupling $\mathcal{L} \supset k (a/f_a) F_C \wedge F_C / (8 \pi^2)$ which yields a potential $V(k a / f_a)$ which is $2\pi$ periodic in its argument. What is the domain wall number in this case? A minimal string requires $\theta_\phi \rightarrow \theta_\phi + 2\pi$, but this is accomplished by a PQ transformation of $\alpha=2\pi/k$, under which the axion does not rotate a full $2\pi$ around its potential and pass $k$ maxima, but rather only rotates $a \rightarrow a + 2\pi f_a/k $ and passes solely one maxima. A string around which $a \rightarrow a + 2\pi f_a$ is in fact a winding $k$ string. The minimal winding string has only one domain wall. 

\paragraph{Lazarides-Shafi} A possible subtlety was pointed out by Lazarides and Shafi \cite{Lazarides:1982tw},  who realized that a na\"{i}ve analysis may misidentify the strings with minimal winding. They realized that if a $\mathbb{Z}_N$ subgroup of the PQ symmetry overlaps with a subgroup of an exact symmetry $G$ then, in our language, this nontrivial `global structure' $(G \times U(1)_{\rm PQ})/\mathbb{Z}_N$ can result in a reduction of the effective domain wall number $N_{\rm DW} \mapsto N_{\rm DW}/N$. In their case this was the $Spin(10)$ GUT and a global symmetry structure $(Spin(10) \times U(1)_{\rm PQ})/\mathbb{Z}_4$.

In our case, it is important to realize that the $\mathbb{Z}_2 \subset U(1)_{\rm PQ}$ acts in the same way as the center $\mathbb{Z}_2 \subset SU(2)_L$ and/or $\mathbb{Z}_2 \subset U(1)_Y$ (see Table \ref{tab:irchargesCubic}).\footnote{In fact there is further subtlety because this theory additionally has a good, anomaly-free symmetry in $U(1)_{Q-N_cL}$. Any combination $\text{PQ} + c (Q - N_cL)$, $c\in\mathbb{Z}$ is also a PQ symmetry with the same anomaly. So in general we could have started with any such combination, and the statement of global structure is that there is some such combination where the PQ symmetry overlaps with the gauge symmetry.\label{foot:PQStructureSubtlety}} This means the symmetry group in this theory is actually 
\beq
\frac{ G_{\rm EW} \times U(1)_{\rm PQ} }{ \mathbb{Z}_2 }, \;\;\;\; G_{\rm EW} = \frac{SU(2)_L \times U(1)_Y}{\Gamma}, \;\; \Gamma = 1 \; {\rm or} \; \mathbb{Z}_2
\eeq
where $\Gamma = \lbrace 1, \mathbb{Z}_2\rbrace$ is the global structure of the electroweak gauge group itself, and the other $\mathbb{Z}_2$ is the overlap of the PQ symmetry with the $\mathbb{Z}_2$ center of either $SU(2)_L$ or $U(1)_Y$.
The mechanism we describe below will work regardless of $\Gamma$.
\begin{figure}
    \centering
    \includegraphics[width=0.5\linewidth]{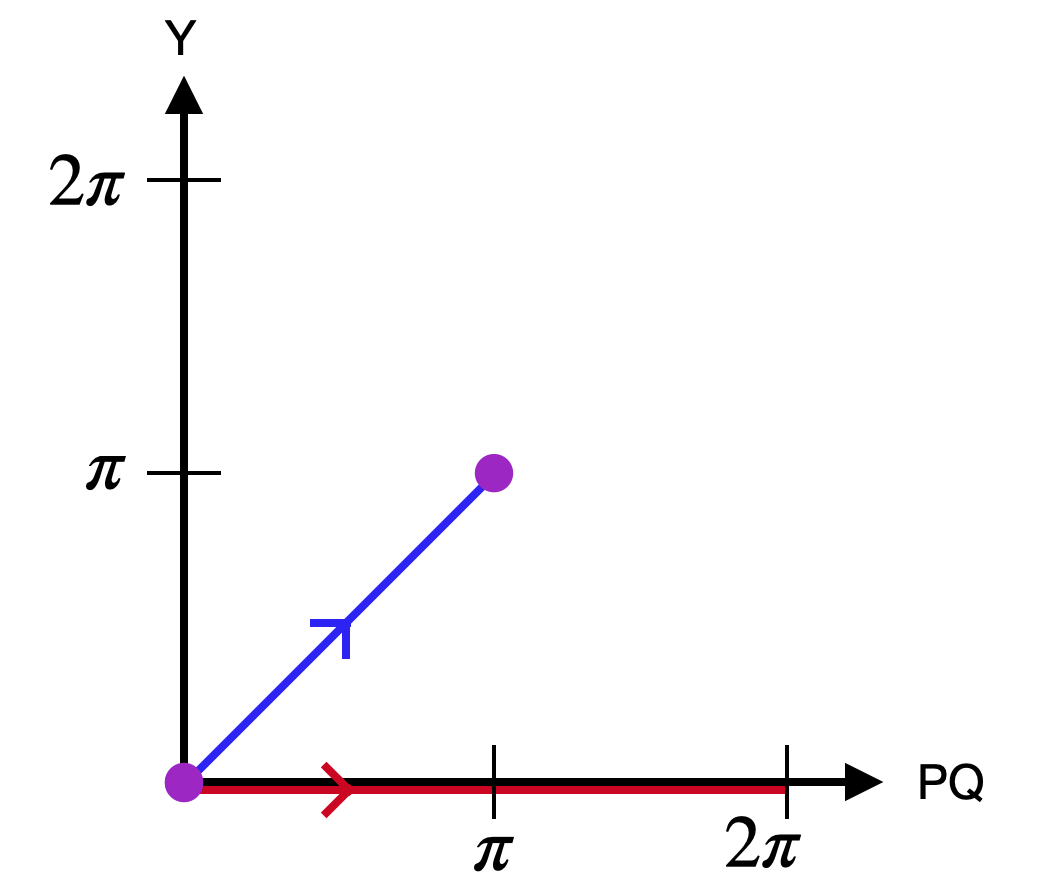}
    \caption{The pure PQ string consists of winding along only the PQ direction (red). On the other hand, non-trivial global structure between PQ and  hypercharge ($Y$) allows more minimal axion string which winds around both the PQ direction and the $Y$ direction by $\pi$ (blue). The quotient $(U(1)_Y \times U(1)_{\rm PQ})/\mathbb{Z}_2$ exactly sets the purple points to be equivalent.} 
    \label{fig:DFSZ_string_winding}
\end{figure}

The Lazarides-Shafi mechanism is fundamentally about properly identifying the spectrum and dynamics of the cosmic string -- domain wall network structure.\footnote{See \cite{Suzuki:2026xvf} for recent study of a TQFT which is postulated to describe the IR of Lazarides-Shafi models.} For our interests we can consider the low-energy theory where the dynamics of the radial modes are integrated out, and we can characterize the cosmic strings as maps from the $\mathbb{S}^1$ on the transverse spatial plane to an $\mathbb{S}^1$ in field space. In particular, string solutions exist when the vacuum manifold has a non-contractible $\mathbb{S}^1$, $\pi_1(\mathcal{M}_{\rm vac}) \neq 1$. 
Such a flat direction in field space should be the result of a symmetry which enforces this degeneracy. 
As we discuss now, the presence of non-trivial global structure between symmetries, here between global and gauged symmetries, has an important effect on the spectrum of cosmic strings.

Let us discuss the implications of $\mathbb{Z}_2$ global structure on the string-domain wall network. We will find that this allows cosmic strings which minimally wind around the axion field space such that $N_g$ domain walls are attached to then, rather than the naive $2 N_g$.

In the case of global structure with hypercharge, the quotient implies that a $\pi$ rotation in the $U(1)_{\rm PQ}$ direction acts the same way as a $\pi$ rotation in the $U(1)_Y$ direction. This means we may write down a cosmic string solution which only rotates halfway around $U(1)_{\rm PQ}$ (see Figure~\ref{fig:DFSZ_string_winding} for an illustration)
\begin{equation}
\vec{\vartheta}(\chi)
= \begin{pmatrix}
        v_u \theta_u(0)  \\
        v_d \theta_d(0)  \\
        v_\phi \theta_\phi(0) 
    \end{pmatrix} 
    + \frac{\chi}{2} \left(\vec{Y} + \vec{X}\right), \qquad 
    \vec{\vartheta}( 2\pi) = \begin{pmatrix}
        v_u \left(\theta_u(0) + 2\pi \right) \\
        v_d \left(\theta_d(0) - 4 \pi \right) \\
        v_\phi \left(\theta_\phi( 0) + 2\pi \right)
    \end{pmatrix} = \vec{\vartheta}(0).
    \label{eq:vecthetachi_1}
\end{equation}
Clearly we indeed get a well-defined result in the sense that all $\theta$ variables wind around integer multiples of $2\pi$. 
As we have explicitly constructed the string solution in terms of the symmetry generators, it is clear that the effect around the cosmic string is a rotation by $\alpha_{\rm PQ} = \pi$ and $\alpha_{Y} = \pi$. This makes it easy to see the effect on the axion by projecting $\vec{\vartheta}$ along $\hat{a}$, that is $a = \vec{\vartheta} \cdot \hat{a}$.

In all, the axion winds by $\Delta a = \pi f_a$ around the string, and the longitudinal mode winds by $\Delta y = 2 \pi \frac{v_u^2 + 2 v_d^2}{v_{\rm EW}}$ due to its additional shift under the Peccei-Quinn symmetry. This fractional winding will lead to fractional magnetic fluxes through these strings.
The fact that the magnetic flux is fractional is consistent with prior results \cite{Abe:2020ure,Niu:2023khv} and of course consistent with the magnetic one-form symmetry, for which quantized magnetic flux is only necessary when integrated over a closed 2-surface. We expect there is more to say elsewhere about these strings with fractional fluxes in the language of generalized symmetries. 

Alternatively to the winding around $U(1)_Y$, if there is shared global structure with $SU(2)_L$ then we can have a string which combines a $\mathbb{Z}_2$ center rotation with a PQ rotation. We recall that it is $T_L^3$, the non-trivial diagonal generator of $SU(2)_L$ which exponentiates to a center rotation: $\exp i \pi T_L^3 \equiv \exp i \pi \mathds{1}_L$. We use a convention where the $T_L^3$ generator has eigenvalues $\pm 1$ for the fundamental, and define again a vector of charges times vevs $\vec{T} = (-v_u,+v_d,0)$. Then we have cosmic string solutions
\begin{equation}
\vec{\vartheta}(\chi)
= \begin{pmatrix}
        v_u \theta_u(0)  \\
        v_d \theta_d(0)  \\
        v_\phi \theta_\phi(0) 
    \end{pmatrix} 
    + \frac{\chi}{2} \left(\vec{T} + \vec{X}\right), \qquad 
    \vec{\vartheta}( 2\pi) = \begin{pmatrix}
        v_u \left(\theta_u(0) - 2\pi \right) \\
        v_d \left(\theta_d(0) + 0 \right) \\
        v_\phi \left(\theta_\phi( 0) + 2\pi \right)
    \end{pmatrix} = \vec{\vartheta}(0).
    \label{eq:vecthetachi_2}
\end{equation}

In general, the global structure allows a family of string solutions in the sector of axion-winding-one sector which are characterized by different windings through the electroweak gauge group. Determining which is the stable string in this topological sector requires further detailed consideration of the full solution, which we leave for future work.

\subsubsection{String Formation}

In a realistic model there is an energy scale hierarchy $v_\phi \gg v_u, v_d$.\footnote{For simplicity we consider $\tan \beta \sim 1$ such that both $H_u, H_d$ get vevs simultaneously. If $v_u\gg v_d$ or $v_d \gg v_u$ there is another stage and the story can be more complex.} We consider briefly how string formation dynamically occurs. At high temperatures $v_{u,d} \lesssim T \lesssim v_\phi$, only the singlet $\phi$ has condensed. Since the axion mode is composed only of pseudoscalar modes of scalars with vevs, the axion direction in field space does not yet have its infrared form. We can make use of the formalism above just setting $v_{u,d} \rightarrow 0$ and we consistently have the axion mode 
\begin{equation}
    a = v_\phi \theta_\phi = \frac{f_a}{2} \theta_\phi.
\end{equation}
The condensation of $\phi$ results in the production of cosmic strings. Thus far the condensed scalars are not charged under electroweak gauge symmetry and so these strings do not have any gauge flux running through their core, but nonetheless the Lazarides-Shafi mechanism is already at play because the quotient dictates that the singlet $\phi$ has charge 2 under $U(1)_{\rm PQ}$. The resulting minimal winding strings come from those above by turning off the electroweak vevs, and we have 
\beq 
a (0) \mapsto a (2\pi) =  v_\phi \left( \theta_\phi (0) + \frac{2\pi}{2} \cdot 2 \right) = a(0) + \pi f_a. 
\label{eq:axion string UV}
\eeq 
Manifestly, while $\theta_\phi \mapsto \theta_\phi + 2\pi$, the proper normalization of the axion field has $\theta_\phi$ transformed by $\alpha_{\rm PQ} = \pi$, again because $\phi$ carries PQ charge 2. 

Now let us consider $T \lesssim v_{u,d}$, below the electroweak symmetry breaking. The pre-existing $\theta_\phi$ strings become dressed with $\theta_u, \theta_d$ phases, as demanded by the cubic term in the potential---any deviation from this would lead to an areal divergence for the string tension $dE/d\ell \sim \lambda_3 v_u v_d v_\phi R^2 \cos(\theta_u + \theta_d + \theta_\phi)$.

\begin{figure}
    \centering
    \includegraphics[width=0.8\linewidth]{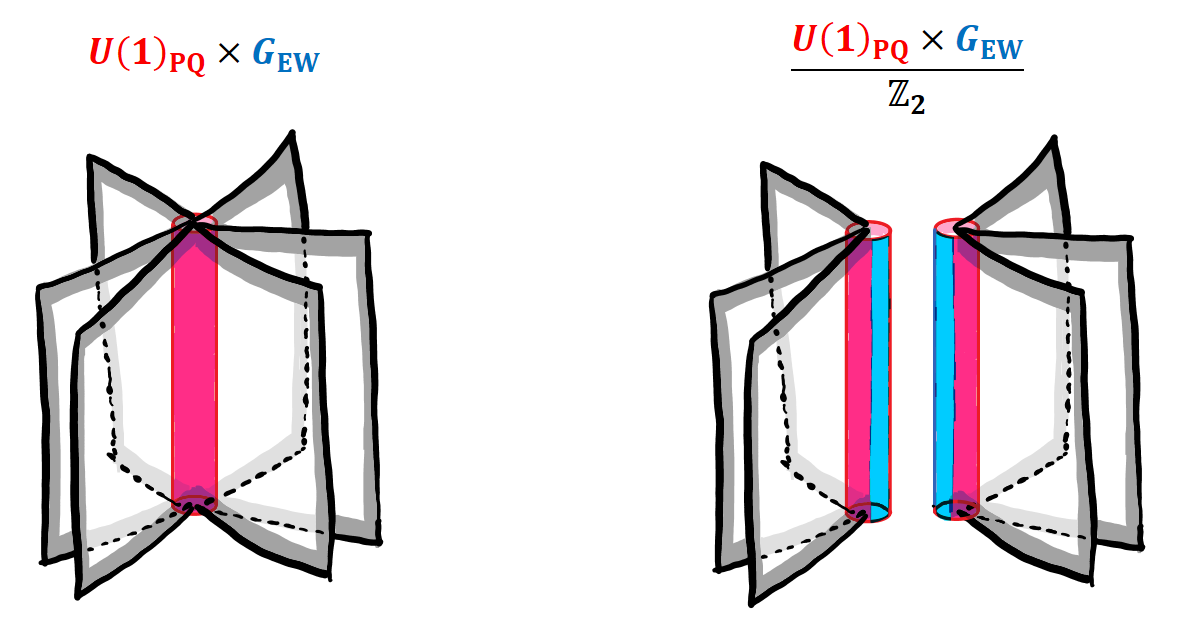}
    \caption{Axion domain wall-string configuration in a theory with $U(1)_{\rm PQ} \times G_{\rm EW}$ (left) vs in a theory with $[ U(1)_{\rm PQ} \times G_{\rm EW}] / \mathbb{Z}_2$ (right). Red color indicates PQ winding and blue color indicates winding along the gauge direction, e.g.~$U(1)_Y$.  }
    \label{fig:DW-string_DFSZ}
\end{figure}

The axion direction is given by $a$ in Eq.~\eqref{eqn:cubicAxMode}, and by construction the path around the string in field space still gives 
\beq 
a (0) \mapsto a (2\pi) = a (0) + \pi f_a
\label{eq:axion string in IR}
\eeq 
because of the facts that $\hat{a}$ is perpendicular to $\hat{Y}$ by construction, and relatedly that $\hat{a}\cdot\vec{X}=f_{a}$. In the full Eq.~\eqref{eq:vecthetachi_1} we see directly that with all three phases included the $2\pi$-winding $\theta_\phi$ configuration is now dressed up with $2\pi n$-winding of $\theta_u, \theta_d$ with $n \in \mathbb{Z}$, which has required that the winding through $U(1)_{\rm PQ}$ is supplemented by winding through the electroweak gauge group. As discussed above, the path around the string by $\alpha_{\rm PQ} = \pi = \frac{2\pi}{2 N_g} \times N_g$ passes only halfway around the axion field space, and there are $N_g$ domain walls attached to the string. The would-be $\mathbb{Z}_2$ factor of the DFSZ domain wall problem is illusory, and has been solved automatically by global structure. See Figure~\ref{fig:DW-string_DFSZ} for illustration.

Finally, we note that in our model we need not confront the subtleties discussed recently by \cite{Lu:2023ayc}, who nicely emphasize that Lazarides-Shafi cannot be considered as a purely infrared solution to the domain wall problem. Gauging $\mathbb{Z}_{N}$ does not merely identify these vacua; it broadens the cosmic string spectrum by allowing $1/{N}$ fractional winding strings which have $N_{\rm DW}/N$ domain walls attached to them. However, the domain wall problem is only resolved if the minimal winding strings not only appear in the spectrum but are in fact produced in the early universe, and the question of which Peccei-Quinn strings will be produced depends on further ultraviolet data. 
Indeed, such a minimal winding string may only appear from a complicated combination of scalar windings in a generic UV theory. However, as we have just shown, our UV model does indeed produce the minimal winding strings.

\subsection{Embedding in Gauge-Gauge Global Structure and Non-Invertible Symmetry} \label{sec:gauge-gauge}

Now that we have identified the precise version of the domain wall problem in the cubic DFSZ model, in this section we propose a novel solution to the remaining $\mathbb{Z}_{N_g}$ domain wall problem based on non-invertible Peccei-Quinn symmetry. 
We will make use of a structure built in to the SM degrees of freedom which allows us to introduce further $\mathbb{Z}_3$ global structure between two gauge symmetry factors, and will lead to $\mathbb{Z}_3$ non-invertible PQ symmetry. The resulting ``non-invertible domain walls'' can be destabilized by embedding this IR theory into a UV theory whose small instanton effects induce explicit breaking of non-invertible symmetry \cite{Cordova:2023her}.\footnote{Non-invertible symmetry breaking via a UV completion and resulting small instanton effects was introduced in \cite{Cordova:2022ieu} and later applied to particle physics \cite{Cordova:2022fhg, Cordova:2023her,  Cordova:2024ypu, Delgado:2024pcv}. }

To this end, we recall that gauging some of the approximate global symmetries of the SM fields is a simple and elegant way to go beyond the SM, as is familiar in the `vertical' case from grand unification of the SM gauge interactions. Upgrading some of the approximate global flavor symmetries to spontaneously broken gauge symmetries is just as well-motivated, and our recent works \cite{Cordova:2024ypu,Delgado:2024pcv} have explored gauging the quark flavor symmetries in a way that intermingles them with color as
\begin{equation}
        \frac{SU(3)_C \times U(1)_{Q_1 + Q_2 - 2Q_3}}{\mathbb{Z}_3} \quad \text{ or } \quad \frac{SU(3)_C \times SU(3)_H}{\mathbb{Z}_3}.
\end{equation}
It is notable that the $\mathbb{Z}_3$ global structure appearing above is intrinsically rooted in the fact that $N_c = N_g = 3$ in the Standard Model.
These theories possess fractional instantons as a result of their non-trivial global structure, and 
this can result in some of the approximate Peccei-Quinn symmetries becoming non-invertible. For details on non-invertible symmetries in 4d QFT, including explicit construction of symmetry defect operators and the case with fractional instantons, can be found in \cite{Hong:2025qbw}.

Let us focus on describing schematically the $SU(3)^2/\mathbb{Z}_3$ theory; for more detail and for the other case we refer to \cite{Cordova:2024ypu}. While the instanton number $\mathcal{N}_C$ is necessarily an integer in an $SU(3)$ Yang-Mills theory, the instanton spectrum in $SU(3)/\mathbb{Z}_3$ becomes richer and includes fractional values $\mathcal{N}_C\in \mathbb{Z}/3$. More explicitly, for $PSU(N) = SU(N) / \mathbb{Z}_N$ it takes the form~\footnote{For more details on fractional instanton, see for e.g.~\cite{Anber:2021iip, Brennan:2023mmt, Cordova:2024ypu, Hong:2025qbw}. } 
\beq
\mathcal{N}_{PSU(N)} =  \frac{N-1}{N} \int \frac{w_2 \wedge w_2}{2} , \;\;\;\;\;\;\;\; \oint_{\Sigma_2} w_2 = 0, 1, \cdots, N-1 \, .
\eeq
The $\mathbb{Z}/ N \mathbb{Z}$-valued 2-form field $w_2$ is known as the second Stiefel--Whitney class. In a theory with a spin structure, which is necessary to introduce fermions, $\int \frac{w_2 \wedge w_2}{2} \in \mathbb{Z}$.

In the $SU(3)^2/\mathbb{Z}_3$ theory the quotient by the diagonal $\mathbb{Z}_3$ interlinks the fractional part of the instanton numbers of the two factors and enforces their equality up to the addition of integers:
\begin{equation}
    \mathcal{N}_C = \mathcal{N}_H \text{ (mod 1)}\,.
\end{equation}
In terms of $\mathbb{Z}_3$-valued $w_2$ fields, this means
\beq
\frac{1}{3} \oint_{\Sigma_2} w_2 (C) = \frac{1}{3} \oint_{\Sigma_2} w_2 (H) \;\; \text{(mod 1)} \;\; \Rightarrow \;\;  w_2 (C) = w_2 (H) + 3 X_2
\eeq
where $w_2 (C)$ and $w_2 (H)$ are the second Stiefel--Whitney classes of color and flavor gauge group, respectively, and  $X_2 \in H^2 (M, \mathbb{Z})$, i.e. $\oint_{\Sigma_2} X_2 \in \mathbb{Z}$ with $\Sigma_2$ a 2-cycle. The above expression should be viewed as a cohomology class.

In the $SU(3)^2/\mathbb{Z}_3$ theory, the anomalous Ward identity now leads to 
\begin{equation}
    \Delta Q_{\rm PQ} = \mathcal{A}_{SU(3)^2_C {\rm PQ}} \mathcal{N}_C + \mathcal{A}_{SU(3)^2_H {\rm PQ}} \mathcal{N}_H,
\end{equation}
where the anomaly coefficients are equal because the quarks transform as $(3,3)$ under $su(3)_C \oplus su(3)_H$ and so the most breaking can be found in a minimal instanton with $\mathcal{N}_C = \mathcal{N}_H = 1/3$ which effects $\Delta Q_{\rm PQ} = 4$. With PQ broken by 6 by integer instantons and 4 by fractional instantons, this means that in the gauged flavor theory with the $\mathbb{Z}_3$ quotient, the Peccei Quinn symmetry is explicitly broken as 
\begin{equation}
    U(1)_{\rm PQ} \rightarrow \mathbb{Z}_2 \ ({\rm invertible}) \times \mathbb{Z}_3 \ ({\rm non \text{-}invertible}).
\end{equation}

As we showed above, the domain wall problem associated with the $\mathbb{Z}_2$ factor is taken care of by $\mathbb{Z}_2$ global-gauge global structure and an analog of Lazarides-Shafi mechanism, and the real domain wall problem for DFSZ is therefore this $\mathbb{Z}_3$ degeneracy. It is important to realize that in our construction the remaining $\mathbb{Z}_3$ symmetry is fully a non-invertible chiral symmetry, and hence the resulting domain wall problem can be solved by the paradigm of non-invertible naturalness: identifying a non-invertible spurion in an IR theory points to a UV mechanism to generate a small number through nonperturbative gauge theory effects. We follow this strategy in our UV completion presented in Section~\ref{sec:CFunification}, which indeed lifts the degeneracy of vacua enforced by this non-invertible symmetry breaking. This means that $\mathbb{Z}_3$ domain walls (attached to a minimal axion string) can be made unstable by the bias potential produced by the symmetry breaking effects and the resulting pressure will collapse the string-domain wall system. 

\subsection{Remark: Abelian Gauge-Global Global Structure and Non-Integer Basis Changes}

\begin{table}\centering
\large
\renewcommand{\arraystretch}{1.3}
\begin{tabular}{|c|c|c|c|c|c|c|}  \hline
 & $Q$ & $\bar u$ & $\bar d$ & $H_u$ & $H_d$ & $\phi$ \\ \hline
$U(1)_{\rm PQ'_3}$ & $0$ & $+2$ & $-1$ & $-2$ & $+1$ & $+1$ \\ \hline
\end{tabular}\caption{Another choice of PQ symmetry basis in the cubic model.}\label{tab:alternativeCubic}
\end{table}

In Section \ref{sec:global-gauge} we showed that global structure between the PQ symmetry and the gauge group can reduce the domain wall number. In the DFSZ model, this global structure can exist between $U(1)_{\rm PQ}$ and the electroweak group of the Standard Model, either or both of $SU(2)_L$ and $U(1)_Y$. As exhibited, in either case there is a minimal winding string which utilizes the $\mathbb{Z}_2$ identification with the gauge group.

In the particular case where the global structure is between $U(1)_{\rm PQ}$ and an \textit{abelian} gauge group, here $U(1)_Y$, we can describe the correct physics in another way that sidesteps a Lazarides-Shafi description by making the minimal strings manifest from the outset. The quotient implies a relation between the charges
\begin{equation}
    G \supset \frac{U(1)_Y \times U(1)_{\rm {PQ}}}{\mathbb{Z}_2} \qquad \Rightarrow \qquad x_i + y_i = 0 \quad (\text{mod} \ 2),
\end{equation}
where $x_i$ is the PQ charge of species $i$ and $y_i$ its hypercharge. This implies that we are allowed to make a non-integer-valued basis change such as
\begin{equation}
    \rm{PQ'} = \frac{1}{2} \left( \rm{PQ} - Y\right),
\end{equation}
and the fields will all have integer charges under $\rm{PQ'}$. 
In terms of our description in Section \ref{sec:whosThatAxion}, this corresponds to a change
\begin{equation}
    \vec{X} \rightarrow \vec{X}' = \half \vec{X} - \half \vec{Y}.
\end{equation}
We showed already above that neither a shift by $\vec{Y}$ nor a re-normalization of the charge modify the identification of the axion mode. Importantly, in this case we do change $\mathcal{A} \rightarrow \mathcal{A}/2$, however since the ${\rm PQ'}$ charges are valued in the integers, a rotation by $\alpha_{\rm PQ'} = 2\pi$ continues to act as the identity so that topological aspects are not obscured. 

Then in the $\rm{PQ'}$ basis, the domain wall number is indeed equal to the anomaly $\mathcal{A} = N_g$ and it is not necessary to speak about a Lazarides-Shafi mechanism. The minimal cosmic string indeed has $a \rightarrow a + 2\pi f_a$, passing through three minima of the axion potential, such that the $\mathbb{Z}_{N_g}$ domain wall problem is manifest.\footnote{Parenthetically, we note that there are yet broader non-integer PQ redefinitions that are allowed as a result of additional global structure \begin{equation}
    \frac{\left(U(1)_{\rm {PQ}_3} \times \left[U(1)_Y\right) \times U(1)_{Q-N_cL}\right]}{\mathbb{Z}_2 \qquad \times \qquad \mathbb{Z}_3}.
\end{equation} which by similar reasoning allows for redefinitions such as
\begin{equation}
    \rm{PQ}''_3 = \frac{1}{2} \left(\rm{PQ}_3 - \frac{1}{3}\left(Y + 2 (Q-N_c L)\right)\right).
\end{equation}
One may have expected $Q-N_cL$ to be irrelevant to the PQ identification since the scalars are uncharged under this, but the global structure nonetheless allows nontrivial PQ redefinitions involving this factor as well. The invariance of the scalar sector under this symmetry makes manifest that this redefinition does not disrupt the axion identification.
}
We note that an important aspect of our analysis given in previous subsections is that even if one works on a basis of symmetry inequivalent to $\vec{X}'$, so long as one properly incorporates the global structure, one is guaranteed to get the right answer. In addition, in general, it is not guaranteed that a theory admits a symmetry basis that ``trivializes'' the global structure.

In this basis we can find a string where the phase rotation is
\begin{equation}
\vec{\vartheta}(\chi)
= \begin{pmatrix}
        v_u \theta_u(0)  \\
        v_d \theta_d(0)  \\
        v_\phi \theta_\phi(0) 
    \end{pmatrix} 
    + \chi \ \vec{X}', \qquad 
    \vec{\vartheta}( 2\pi) = \begin{pmatrix}
        v_u \left(\theta_u(0) - 4\pi \right) \\
        v_d \left(\theta_d(0) + 2 \pi \right) \\
        v_\phi \left(\theta_\phi( 0) + 2\pi \right)
    \end{pmatrix} = \vec{\vartheta}(0).
    \label{eq:vecthetachi_3}
\end{equation}
As dictated by the construction, now this minimal winding string has the effect of a rotation by $\alpha_{\rm PQ} = 2\pi$ with no associated rotation by hypercharge necessary. 

Preparing for the quartic case below, we note that this PQ redefinition seems to be available only when the global structure can be understood as solely involving a continuous abelian gauge group. 
However, as we noted above, the Lazarides-Shafi mechanism can work when PQ overlaps with abelian and/or nonabelian gauge groups. 
In particular in the cubic theory, if we add a field with $x_i = 0, y_i = 1$ which is an $SU(2)_L$ singlet, this $\rm{PQ'}$ basis will no longer be integer-valued.\footnote{As emphasized in Footnote \ref{foot:u1charge}, the problem here with the ${\rm PQ'}$ charges now being half-integral is that a $U(1)_{\rm PQ'}$ transformation by $2\pi$ is no longer the identity, but rather only $4\pi$. This means that still the minimal cosmic string would still involve both a $U(1)_{\rm PQ'}$ transformation and a $U(1)_Y$ transformation.} 
While there is no longer global structure between PQ and hypercharge, we will still have the global structure $(SU(2)_L \times U(1)_{\rm PQ})/\mathbb{Z}_2$, and the domain wall number will still be reduced to $N_g$. The Lazarides-Shafi mechanism will still be active, and this time there is no change of basis which trivializes it. We will find ourselves in just this situation in the quartic case below. 

\section{Quartic DFSZ} \label{sec:quartic}

The usual quartic model has a complex scalar $S = \rho e^{is}$ interacting with the 2HDM through the marginal quartic term, which in the linear and nonlinearly realized phases looks like
\begin{align}
    E > \langle \rho \rangle:& \ \mathcal{L} \supset \lambda_4 H_u H_d S^2 + \text{h.c.}, \label{eq:quarticAbove} \\
    E < \langle \rho \rangle:& \ \mathcal{L} \supset \lambda_4 H_u H_d \langle \rho \rangle^2 e^{i 2s} + \text{h.c.}, \label{eq:quarticBelow}
\end{align}
and one choice of PQ charges is as in Table~\ref{tab:alternativeQuartic}. Notice that the singlet $S$ has unit PQ charge now, so there exists no non-trivial overlap between $U(1)_{\rm PQ}$ and SM gauge symmetry. This implies that the $\mathbb{Z}_2$ Lazarides-Shafi mechanism is no longer active. Since the PQ charges of the fermions are the same as previously, this implies we here have a full $\mathbb{Z}_{2N_g}$ domain wall problem.\footnote{Parenthetically, we remark that the original DFS paper \cite{Dine:1981rt} analyzes the quartic model and at first glance would seem to imply $N_{\rm DW} = N_g$, as the QCD anomaly is found to be $N=N_g$ (their Eq.~19). However, they have assigned the PQ charge of $S$ to be $1/2$ (their Eq.~4), such that only a PQ rotation by $4\pi$ acts as the identity (recall our warning in Footnote \ref{foot:u1charge}), meaning their analysis indeed agrees $N_{\rm DW} = 2N_g$.} 

Our strategy to solve this $\mathbb{Z}_{2N_g}$ domain wall problem is as follows. First, inspired by the role of $\mathbb{Z}_2$ global structure between $U(1)_{\rm PQ}$ and SM gauge group in reducing the domain wall problem to $\mathbb{Z}_{N_g}$, we construct a left-right extension of the theory where a new gauged $SU(2)_R$ introduces $\mathbb{Z}_2$ global structure, reducing the domain wall spectrum from $\mathbb{Z}_{2N_g}$ to $\mathbb{Z}_{N_g}$ in the form of Lazarides-Shafi mechanism. This then sets the stage for fully solving the problem based on non-invertible symmetry breaking, in which $\mathbb{Z}_3$ quotient between two gauge group factors is the key.

\begin{table}\centering
\large
\renewcommand{\arraystretch}{1.3}
\begin{tabular}{|c|c|c|c|c|c|c|}  \hline
 & $Q$ & $\bar u$ & $\bar d$ & $H_u$ & $H_d$ & $S$ \\ \hline
$U(1)_{\rm PQ}$ & $+1$ & $0$ & $0$ & $-1$ & $-1$ & $+1$ \\ \hline
\end{tabular}\caption{One choice of PQ symmetry basis in the usual low-energy quartic DFSZ model. }\label{tab:alternativeQuartic}
\end{table}

\subsection{$\mathbb{Z}_2$ Global Structure from $SU(2)_R$ Embedding} \label{sec:su2rEmbedding}

\begin{table}\centering
\large
\renewcommand{\arraystretch}{1.3}
\begin{tabular}{|c|c|c|c|c|c|}  \hline
 & $SU(2)_L$ & $SU(2)_R$ & $U(1)_{Q-N_cL}$ & $U(1)_{\tilde Q}$ & $U(1)_{\rm PQ}$ \\ \hline

$Q$ & $2$ & $-$ & $+1$ & $+1$ & $0$ \\ \hline

$\bar Q = \begin{pmatrix}
    \bar u \\ \bar d
\end{pmatrix}$ & $-$ & $2$ & $-1$ & $-1$ & $+1$ \\ \hline

$H = \begin{pmatrix}
    H_d \\ H_u
\end{pmatrix}$ & $2$ & $2$ & $0$ & $0$ & $-1$ \\ \hline

$\varphi$ & $-$ & $2$ & $-3$ & $0$ & $+1$ \\ \hline
$\varphi'$ & $-$ & $2$ & $+3$ & $0$ & $+1$ \\ \hline

\end{tabular}\caption{Charges of the fields under the gauge and global symmetries of the left-right DFSZ model. $U(1)_{Q-N_c L}$ is gauged, $U(1)_{\tilde{Q}}$ is global and anomaly-free, $U(1)_{\rm PQ}$ is anomalous.
}
\label{tab:irchargesQuartic}
\end{table}

We will make use of the same mechanism by enlarging our invisible axion sector to bear out a `right-handed' version of the PQWW axion. To this end, we introduce two PQ-charge carrying scalars, $\varphi, \varphi'$, which are also charged under the upgraded gauged symmetry $SU(2)_R \times U(1)_{Q-N_cL}$ (see Table~\ref{tab:irchargesQuartic}). In this model, the axion will dominantly be composed of the pseudoscalar modes of $\varphi, \varphi'$ charged under $SU(2)_R\times U(1)_{Q-N_cL}$ (2$\varphi$DM) in just the same way that the ``visible PQWW axion'' comes from a 2HDM charged under $SU(2)_L \times U(1)_Y$. The $\mathbb{Z}_2$ domain wall issue of concern will now be alleviated by the Lazarides-Shafi mechanism in a right-handed version of the mechanism we saw above. The charge assignment is in Table \ref{tab:irchargesQuartic} and we describe a model that realizes this. A crucial feature to mention is that now there exists $\mathbb{Z}_2$ overlap between $SU(2)_R$ and $U(1)_{\rm PQ}$, which will effectively reduce the domain wall number by half, as desired. 

The full story is slightly more subtle. First of all, the electroweak gauge group takes the form
\beq
G_{\rm LR} = \frac{SU(2)_L \times SU(2)_R \times U(1)_{Q-N_cL}}{\Gamma_{\rm LR}}, \;\;\;\; \Gamma_{\rm LR} = \lbrace \mathbb{I}, \; \mathbb{Z}_2 \rbrace
\eeq 
namely, the left-right symmetric DFSZ quartic model is compatible with either choice of global structure. Note that the $\mathbb{Z}_2$ global structure here is crucially different from that possibility in the electroweak gauge group, in that it acts simultaneously among all three gauge factors. 
The charge assignments are such that each field transforms under two factors of the $\mathbb{Z}_2$ center transformation and so these always cancel out. 

The global $U(1)_{\rm PQ}$ is realized with a global structure $\frac{G_{\rm LR} \times U(1)_{\rm PQ}}{\mathbb{Z}_2}$. 
In the basis of Table~\ref{tab:irchargesQuartic}, this $\mathbb{Z}_2$ global structure is realized as $(SU(2)_R \times U(1)_{\rm PQ}) / \mathbb{Z}_2$. If we redefine our PQ charge as ${\rm PQ'} = {\rm PQ} + (Q-N_cL)$, then this time $\mathbb{Z}_2$ quotient appears as $SU(2)_L \times U(1)_{\rm PQ'} / \mathbb{Z}_2$. The crucial point is that there always exists a $\mathbb{Z}_2$ global structure between PQ and electroweak gauge group, here $G_{\rm LR}$.\footnote{As in Footnote \ref{foot:PQStructureSubtlety} there is again a further subtlety due to the presence of a good, anomaly-free global symmetry $U(1)_{\tilde{Q}}$. Having started with an arbitrary combination $\text{PQ} + c \tilde{Q}$, which may not itself possess this global structure, the statement is that there is some such combination which does.}

Below we sketch the model and show the axion physics works as we intend, but we do not here worry about the full details of the Higgses or yukawas, which have a large literature in the context of left-right models (see for example, \cite{Berezhiani:1983hm,Davidson:1987mh,Rajpoot:1987fca,Balakrishna:1987qd,Ma:1989tz,Duka:1999uc,Holthausen:2009uc} and references thereto) and we leave further details to future investigations.

The yukawa sector in this theory can be written more compactly with the left-right bidoublet Higgs\footnote{In an EFT approach we can supplement Eq.~\ref{eq:su2RY} with higher dimension gauge invariant operators such as $\mathcal{L}\supset c_{u}(Q^{a}H_{a\alpha}\varphi_{\gamma}\varphi'_{\beta}\bar{Q}_{\delta}\epsilon^{\alpha\gamma}\epsilon^{\delta\beta})/\Lambda^{2}+c_{d}(Q^{a}H_{a\alpha}\varphi'_{\gamma}\varphi_{\delta}\bar{Q}_{\beta}\epsilon^{\alpha\gamma}\epsilon^{\delta\beta})/\Lambda^{2}$. After condensation of $\varphi$ and $\varphi'$ as in Eq.~\ref{eq:vevvarphi}, the difference between $c_{u}$ and $c_{d}$ may explain the different $y$ for up and down sector operators. Here, $c_{u,d}$ are dimensionless Wilson coefficients and $\Lambda$ is a UV scale below which the operators show up after integrating out high momentum modes.}
\begin{equation}\label{eq:su2RY}
    \mathcal{L} = y Q^a H_{a\alpha} \bar Q^{\alpha},  
\end{equation}
where $a$ is an $SU(2)_L$ index and $\alpha$ is an $SU(2)_R$ index.
In order that the heavy scalars $\varphi, \varphi'$ be charged under the SM PQ symmetry (hence, there is only one PQ symmetry shared among $\varphi, \varphi'$ and SM fields) we must give them some interactions with the Higgses. One allowed gauge-invariant interaction which will match on to the quartic DFSZ model at low energies in \eqref{eq:quarticAbove} is to include
\begin{equation}\label{eq:leftrightPotential}
    V\left(H,\varphi, \varphi' \right) = \lambda_4 \varepsilon^{ab} H_{a\alpha} H_{b\beta} \varphi_\gamma \varphi'_\delta \varepsilon^{\alpha \gamma} \varepsilon^{\beta \delta} + \tilde{V}(H,\varphi, \varphi'),
\end{equation}
which leads to the PQ charges in Table \ref{tab:irchargesQuartic} and where we have shunted to $\tilde{V}$ other interactions which preserve the same symmetries, e.g. $|H|^4, |H|^2 |\varphi|^2$, etc. In fact, it turns out this is the unique interaction consistent with this symmetry direction, and this in turn means that the potential phase in $\lambda_4$ is not a physical parameter.\footnote{The choice of this quartic interaction is to single out the PQ symmetry direction. It is important to note there is only one quartic consistent with this symmetry direction, as otherwise there could be a physical complex phase which would misalign the UV and IR axion potentials. A general tensor $T^{ab\alpha\beta\rho\sigma}$ to contract these fields must be symmetric under the simultaneous exchange of both $a \leftrightarrow b$ and $\alpha \leftrightarrow \beta$. The $SU(2)_L$ indices must be contracted by $\epsilon^{ab}$, which is antisymmetric, so $T^{ab\alpha\beta\rho\sigma} = \epsilon^{ab} T^{\alpha\beta\rho\sigma}$ must also be antisymmetric in $\alpha \leftrightarrow \beta$. This can be $T^{\alpha \beta\rho \sigma} = \epsilon^{\alpha\beta} \epsilon^{\rho\sigma}$ or $T^{\alpha\beta\rho \sigma} = \epsilon^{\alpha [\rho} \epsilon^{\beta \sigma]} = \frac{1}{2} \left( \epsilon^{\alpha \rho} \epsilon^{\beta \sigma} - \epsilon^{\alpha \sigma} \epsilon^{\beta \rho} \right)$. But in 2d, any totally antisymmetric object with 3 or more indices vanishes, e.g.~$A^{[\alpha \beta \rho]}=0$. This implies identities among Levi-Civita symbols, $\epsilon^{\alpha \beta} \epsilon^{\rho \sigma} = \epsilon^{\alpha \rho} \epsilon^{\beta \sigma} - \epsilon^{\alpha \sigma} \epsilon^{\beta \rho}$. Thus there is only one possible quartic, and the phase of the coupling is unphysical because this term breaks a would-be $U(1)$ symmetry generator, as in the cubic model.}

The condensation of $\varphi$ and $\varphi'$ at the scale $v_R^2 = v_\varphi^2 + v_{\varphi'}^2$ together break $SU(2)_R \times U(1)_{Q-N_cL} \rightarrow U(1)_Y$ with $Y = (Q-N_cL) - 6 T^3_R$, as well as breaking $U(1)_{\rm PQ}$, and we can represent their hypercharge-neutral pseudoscalar modes nonlinearly as 
\begin{equation}
    \varphi = \begin{pmatrix}
        0 \\ \frac{v_\varphi}{\sqrt{2}} 
    \end{pmatrix} e^{i \theta_\varphi}, \qquad \varphi' = \begin{pmatrix}
        \frac{v_{\varphi'}}{\sqrt{2}}   \\ 0
    \end{pmatrix} e^{i \theta_{\varphi'}}.
    \label{eq:vevvarphi}
\end{equation}
We see that in this $SU(2)_R$ quartic DFSZ model, the invisible axion sector has been upgraded to match the visible PQWW sector, wherein $H_u$ and $H_d$ break $SU(2)_L \times U(1)_Y \rightarrow U(1)_{\rm EM}$.  
After $SU(2)_R$ is spontaneously broken and the radial modes are integrated out, the potential of Eq.~\eqref{eq:leftrightPotential} reduces to 
\begin{equation}
    V\left(H,\varphi\right) \supset \lambda_4 v_{\varphi}v_{\varphi'} H_u H_d e^{i 2\phi}
    \label{eq:VHvarphi}
\end{equation}
where $2\phi\equiv\theta_{\varphi}+\theta_{\varphi'}$.
This is exactly the desired quartic potential term (in nonlinear representation) suitable for the quartic DFSZ model (see \eqref{eq:quarticBelow}). 

Now let us make sure we can account for all four modes in the infrared once both the left and right sectors have condensed.
One combination proportional to $\theta_u + \theta_d + \theta_\varphi + \theta_{\varphi'}$ gets a large mass due to the explicit breaking from $\lambda_4$. 
We again use $\vec{\vartheta} = (v_u \theta_u, v_d \theta_d,  v_\varphi \theta_\varphi, v_{\varphi'} \theta_{\varphi'})$ as the vector of vevs times angular modes, and $\hat Y$ as the unit vector of vevs times hypercharges, in addition now to $\hat B$ as the unit vector of vevs times $(Q-N_cL)$ charges. The modes eaten by the $(Q-N_cL)$ gauge boson and $Z$ boson, respectively, are 
\begin{equation}
     b = \vec{\vartheta} \cdot \hat{B} = \frac{1}{v_R} (v_{\varphi'}^2 \theta_{\varphi'} - v_{\varphi}^2 \theta_{\varphi}), \qquad y = \vec{\vartheta} \cdot \hat{Y} = \frac{1}{v_{\rm EW}} (v_u^2 \theta_u - v_d^2 \theta_d),
\end{equation}
where $f_b^2 = |\vec{B}|^2=9v_{R}^{2}$ and $f_Y^2 = |\vec{Y}|^2=9v_{\rm EW}^{2}$. 

The axion mode must now be neutral under both $Y$ and $(Q-N_cL)$, which in this case is not difficult to arrange because the phase modes each transform under only one of the two (that is, $\vec{B} \cdot \vec{Y} = 0$), and we can subtract off both components independently 
\begin{equation} \label{eqn:quarticAxMode}
    \hat{a} = \frac{1}{f_a} \left( \vec{X} - (\vec{X}\cdot \hat{B}) \hat{B} - (\vec{X}\cdot \hat{Y}) \hat{Y}\right), 
\end{equation}
where $f_a$ is again defined such that $\hat{a} \cdot \hat{a} = 1$. It is easy to see that this expression is invariant under shifts of $\vec{X}$ by multiples of either $\vec{B}$ or $\vec{Y}$, and that $a = \vec{\vartheta} \cdot \hat{a} \rightarrow a + \alpha_{\rm PQ} f_a$ as is appropriate. Under a transformation $\vec{\vartheta} \rightarrow \vec{\vartheta} + \alpha_Y \vec{Y} + \alpha_B \vec{B} +  \alpha_{\rm PQ} \vec{X}$, we have $y \rightarrow y + \alpha_Y f_Y + \alpha_{\rm PQ} \vec{X} \cdot \hat{Y}$, and $b \rightarrow b + \alpha_B f_b + \alpha_{\rm PQ} \vec{X} \cdot \hat{B}$. Now axion strings will contain magnetic fluxes of both gauge groups. 
Finally, the massive mode is in the direction $\hat{A} \propto \left(\frac{1}{v_u}, \frac{1}{v_d}, \frac{1}{v_\phi}, \frac{1}{v_\phi'}\right)$, and is again orthogonal to the other modes by virtue of the $\lambda_4$ operator respecting the three $U(1)$ symmetries. 

We next discuss allowed cosmic strings of this theory. Similarly to the cubic model, we again find that there is a cosmic string which rotates only by $\pi$ in the PQ direction, the rest of the rotation being in the gauge direction. Considering again a transverse plane with angular coordinate $\chi \in [0,2\pi)$, in the basis of Table~\ref{tab:irchargesQuartic}, we can compensate the rotation of PQ by a rotation along the non-trivial diagonal generator $T^3_R$ of $SU(2)_R$.\footnote{As we discussed above, one is free to work in any consistent basis, and the PQ rotation by $\pi$ can always be completed by a $\pi$ rotation along $G_{\rm LR}$.} We define a vector of $T_R^3$ charges times vevs $\vec{R}= (-v_u,+v_d,-v_\varphi, +v_{\varphi'})$. Then we have
\begin{equation}
    \vec{\vartheta}(\chi) = \vec{\vartheta}( 0) + \frac{\chi}{2} \left(\vec{R} +\vec{X}\right), \qquad \vec{\vartheta}(2\pi) = \begin{pmatrix}
        v_u (\theta_u(0) - 2\pi) \\
        v_d (\theta_d(0) + 0) \\
        v_\varphi (\theta_\varphi(0) + 0) \\
        v_{\varphi'} (\theta_{\varphi'}(0) + 2\pi)
    \end{pmatrix} = \vec{\vartheta}(0),
\end{equation}
and we see that the phase modes smoothly return back to the original point forming a non-trivial winding. This corresponds to the `right-handed' version of the Lazarides-Shafi mechanism compared to the built-in `left-handed' mechanism we showed exists in the cubic model. Importantly, this means that the domain wall number is again reduced by half compared to the naive calculation with a string around which the axion winds by $2\pi$.

\subsection{Anomaly and Non-Invertible Symmetry}

As far as the SM fermions are concerned, this PQ symmetry is the same but with left and right switched compared to the cubic model. This does not affect the anomaly, because QCD is anyway vector-like. 

The ABJ anomaly coefficient for the PQ symmetry defined in Table \ref{tab:irchargesQuartic} in an $SU(3)_C$ instanton background can be easily computed as
\begin{equation}
    \mathcal{A}_{4} = 2 N_g,
\end{equation}
leading to an anomalous conservation equation for the $U(1)_{\rm PQ}$ current 
\begin{align}
    \partial_\mu J^\mu_{\rm PQ} &= \frac{\mathcal{A}_{4}}{16\pi^2} \text{Tr }F_C \tilde{F}_C \\
    \Delta Q_{\rm PQ} &= \mathcal{A}_{4} \mathcal{N}_C
\end{align}
where $F_C$ is the color field strength, $\tilde{F}_C$ is its Hodge dual, and $\mathcal{N}_C = \frac{1}{16\pi^2} \int \text{Tr }F_C \tilde{F}_C$ is the instanton number for a given $SU(3)_C$ configuration. 
This means the ABJ anomaly explicitly breaks $U(1)_{\rm PQ} \rightarrow \mathbb{Z}_{2 N_g}$. This ABJ anomaly data is matched by the Chern-Simons interaction term between the axion and $SU(3)_C$ gauge fields in the IR (but above the QCD confinement)
\begin{equation}
    \mathcal{L} \supset 2 N_g \frac{a}{f_a} \frac{\text{Tr }F_C \tilde{F}_C}{16\pi^2}, 
\end{equation}
which is responsible for generating a potential for the axion as a result of the explicit breaking of $U(1)_{\rm PQ}$ due to the anomaly
\begin{equation}
    V(a) = \Lambda_{\rm QCD}^4 \left[1 - \cos\left(2 N_g \frac{a}{f_a}\right)\right]. 
\end{equation}
As in the cubic case, the domain wall number here is equal to half of the anomaly coefficient due to the $\mathbb{Z}_2$ global structure. To summarize, the extension of the quartic DFSZ model into a left-right version, in which $SU(2)_R \times U(1)_{Q-N_c L}$ is gauged and PQ scalars $\varphi$ and $\varphi'$ are promoted to doublets of $SU(2)_R$ and also carrying $U(1)_{Q-N_c L}$, introduces $\mathbb{Z}_2$ global structure between $U(1)_{\rm PQ}$ and electroweak gauge group. This then resulted in the reduction of the domain wall number to half and hence the domain wall problem. The remaining $\mathbb{Z}_{N_g}$ domain wall problem can be solved by means of non-invertible symmetry and its breaking, as introduced earlier in cubic model.

\section{Embedding in Quark Color-Flavor Unification}
\label{sec:CFunification}

At the level of exhibiting a simple UV theory which accomplishes non-invertible symmetry breaking, our task is relatively straightforward. We upgrade $SU(3)^2/\mathbb{Z}_3 \subset SU(9)$ and the bifundamental quarks become $SU(9)$ fundamentals.

We will exhibit the reduced ultraviolet anomaly coefficients, showing that UV instantons indeed perform non-invertible symmetry breaking. For more detail on spontaneously breaking these theories to get the Standard Model in the infrared, especially the important issue of generating the yukawas, see our earlier \cite{Cordova:2024ypu}.

\subsection{Cubic Model}

\begin{table}\centering
\large
\renewcommand{\arraystretch}{1.3}
\begin{tabular}{|c|c|c|c|c|c|}  \hline
 & $SU(9)$ & $SU(2)_L$ & $U(1)_Y$ & $U(1)_{Q-N_cL}$ & $U(1)_{\rm PQ}$ \\ \hline

${Q}$ & $9$ & $2$ & $+1$ & $+1$ & $+1$  \\ \hline

${\bar u}$ & $\bar 9$ & $-$ & $-4$ & $-1$ & $0$ \\ \hline

${\bar d}$ & $\bar 9$ & $-$ & $+2$ & $-1$ & $0$ \\ \hline \hline

$H_u$ & $-$& $2$ & $-3$ & $0$ & $-1$ \\ \hline

$H_d$ & $-$& $2$ & $+3$ & $0$ & $-1$ \\ \hline

$\phi$ & $-$ & $-$ & $-$ & $0$ & $+2$ \\ \hline

\end{tabular}
\caption{The cubic DFSZ model embedded into quark color-flavor unification. See \cite{Cordova:2024ypu} for a fuller discussion of the UV fields needed in this unified theory to break down to the SM. }\label{tab:cubicUV}
\end{table}

\begin{figure}
\centering
\begin{subfigure}{.5\textwidth}
  \centering
  \includegraphics[width=.7\linewidth]{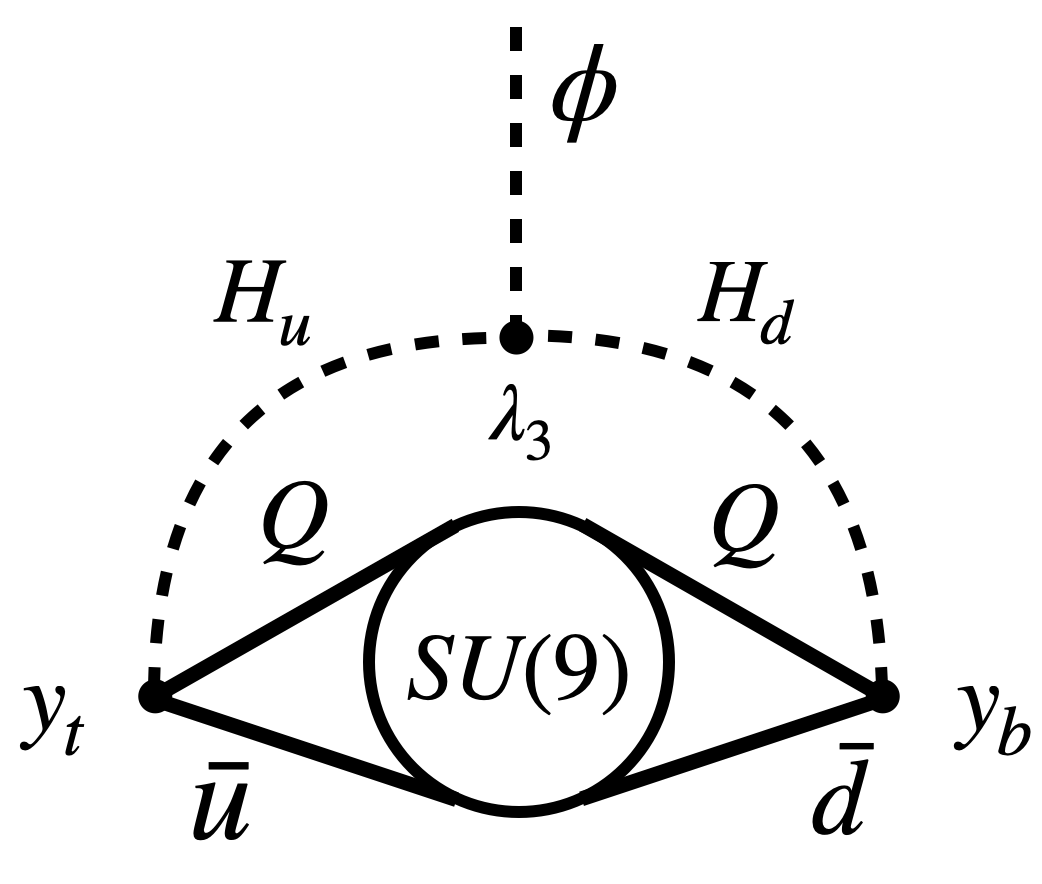}
  \caption{Cubic model.}
  \label{fig:cubicInst}
\end{subfigure}%
\begin{subfigure}{.5\textwidth}
  \centering
  \includegraphics[width=.7\linewidth]{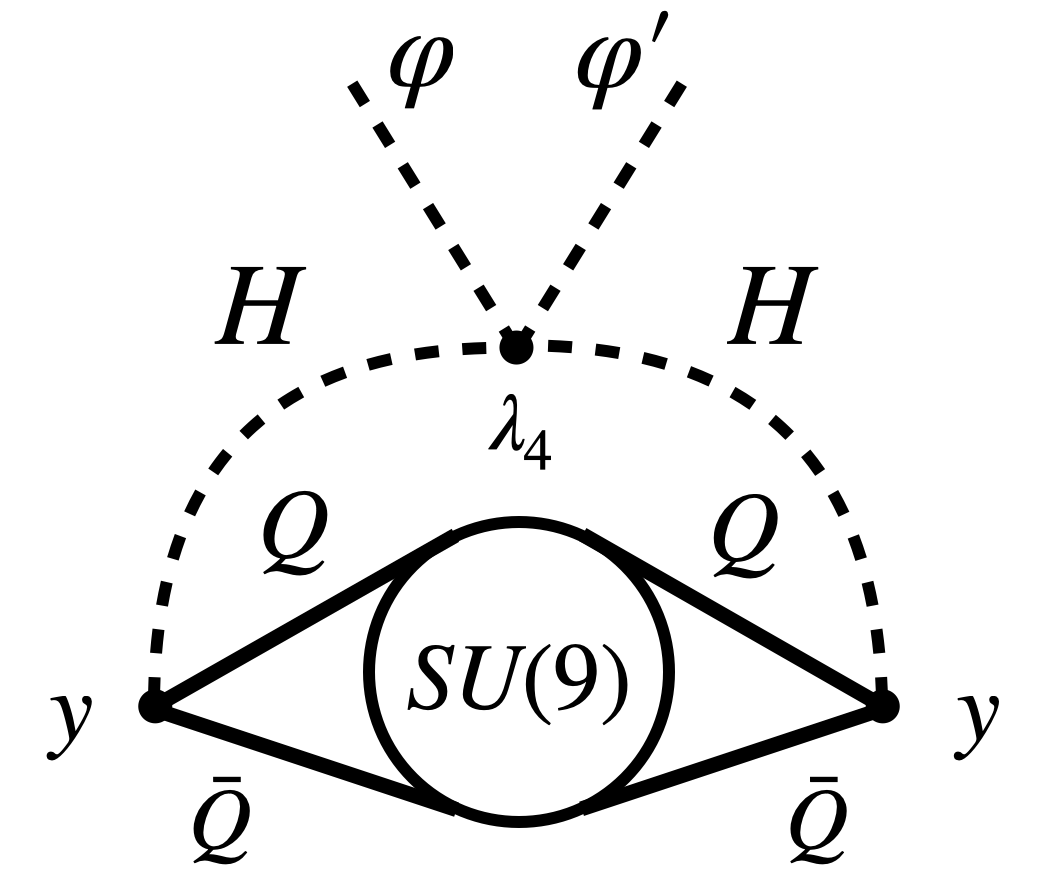}
  \caption{Quartic model.}
  \label{fig:quarticInst}
\end{subfigure}
\caption{One diagram in each $SU(9)$ UV completion generating $\mathbb{Z}_3$ non-invertible PQ symmetry violation. In each case the diagram with $\lambda$ removed also exists.}
\label{fig:SU9Insts}
\end{figure}

We exhibit the symmetries and representations in this UV theory in Table \ref{tab:cubicUV}. The Peccei-Quinn symmetry now has anomaly coefficient 
\begin{equation}
    \mathcal{A}_{SU(9)^2 {\rm PQ}} = 2,
\end{equation}
as we understood would happen from the bottom-up using our non-invertible symmetry analysis. From the UV perspective, the breaking of the IR non-invertible symmetry is from $SU(9)$ instantons. PQ-violating operators are generated from 't Hooft vertices, and the axion potential comes from diagrams where the 't Hooft vertex is connected to the scalar sector, as in the diagram of Figure \ref{fig:cubicInst}. See \cite{Csaki:2023ziz} for systematic NDA for such instanton effects, and \cite{Sesma:2024tcd} for more detailed computation. 

As a result of the anomaly, the instantons associated with $\pi_3 (SU(9) / [ SU(3)^2 / \mathbb{Z}_3])$, often called small instantons or constrained instantons~\cite{tHooft:1976snw,Affleck:1980mp}, generate effective operators which violate the PQ symmetry by 2, including a term mixing $H_u, H_d$ and a tadpole for the singlet scalar, schematically 
\begin{equation}
    V_{\text{small instantons}} = y_t y_b e^{-2\pi/\alpha} v_9^2 H_u H_d + \frac{1}{(4\pi)^2} y_t y_b e^{-2\pi/\alpha} \lambda_3 v_9^2 \phi + \dots
    \label{eq:Vsmallins}
\end{equation}
where $\alpha=g^{2}/(4\pi)$ evaluated at $\rho=(2\pi v_{9})^{-1}$ with $g$ the gauge coupling of $SU(9)$. Here, both lead to an additional potential for the axion. These terms break the non-invertible symmetry of the infrared and so supply the missing harmonics for the axion potential. One of these terms may dominate depending on $\lambda_3 v_\phi / v_u v_d$, but in general \footnote{Note that since $y_t, y_b, \lambda_3$ each break separate $U(1)$ global symmetries, by a standard spurion analysis (see e.g. \cite{Grossman:2017thq}) the phases of these parameters are unphysical and do not cause a potential misalignment of the bias potential. We refer to our earlier \cite{Cordova:2024ypu} for a strategy in this UV theory to produce the CKM phase upon spontaneous symmetry breaking while protecting the axion phase.}

\begin{equation}
	V(a) \sim  y_t y_b e^{-\frac{2\pi}{\alpha}} v_9^2 \left(\frac{\lambda_3  v_\phi}{(4\pi)^2} + \frac{v_{\rm EW}^2 \sin 2\beta}{4}\right)  \left(1 - \cos\left(2\frac{a}{f_a}\right)\right) + \Lambda_{\rm QCD}^4 \left(1 - \cos\left(6\frac{a}{f_a}\right)\right),
\label{eq:axionVcubic}
\end{equation}
where the axion mode has been given in terms of the various pseudoscalar modes above in Eq.~\eqref{eqn:cubicAxMode}. As portended in our analysis in that section, the $\mathbb{Z}_3$ degeneracy has been lifted by the small instantons, while the $\mathbb{Z}_2$ degeneracy does not lead to a domain wall problem due to the Lazarides-Shafi mechanism of $SU(2)_L$.

We can define an effective scale of the small instanton potential $\Lambda_9^4$ as the coefficient of the lower harmonic. It is conceivable that $\Lambda_9 \gg \Lambda_{\rm QCD}$ if $\alpha$ is large, in which case the small instanton effects may dominate the mass of the axion, providing a `heavy axion' candidate. In the other limit, these bumps are a perturbation on top of the QCD potential and we have the standard prediction for the mass of the DFSZ axion, but the string-domain wall network formed during the QCD phase transition will eventually collapse due to the pressure from this potential difference. 
In Section \ref{sec:GW} we will consider this limit. 

\subsection{Quartic Model}

\begin{table}\centering
\large
\renewcommand{\arraystretch}{1.3}
\begin{tabular}{|c|c|c|c|c|c|c|}  \hline
 & $SU(9)$ & $SU(2)_L$ & $SU(2)_R$& $U(1)_{Q-N_c L}$ &$U(1)_{\tilde{Q}}$& $U(1)_{\rm PQ}$ \\ \hline

${Q}$ & $9$ & $2$ & $-$ & $+1$ &$+1$& $0$  \\ \hline

$\bar Q$ & $\bar 9$ & $-$ & $2$ & $-1$ & $-1$ & $+1$ \\ \hline \hline

$H$ & $-$& $2$ & $2$ & $0$ & $0$ & $-1$ \\ \hline

$\varphi$ & $-$ & $-$ & $2$ & $-3$ &$0$ & $+1$ \\ \hline

$\varphi'$ & $-$ & $-$ & $2$ & $+3$ &$0$ & $+1$ \\ \hline

\end{tabular}
\caption{The quartic DFSZ model embedded into the left-right model of Section \ref{sec:su2rEmbedding} then into quark color-flavor unification.}
\label{tab:quarticUV}
\end{table}

The $SU(9)$ embedding proceeds as in the above case, with the charges now as in Table \ref{tab:quarticUV}. The anomaly coefficient is again
\begin{equation}
    \mathcal{A}_{SU(9)^2 \rm PQ} =2
\end{equation}
but we now have that the $\mathbb{Z}_2$ subgroup of PQ is embedded in the $SU(2)_R$ gauge symmetry, and so again does not lead to domain walls.

Now the 't Hooft vertex leads to terms violating PQ by 2 units as in Figure \ref{fig:quarticInst}, giving schematically 
\begin{equation}
    V_{\text{small instantons}} = y_t y_b e^{-2\pi/\alpha} v_9^2 H_u H_d + \frac{1}{(4\pi)^2} y_t y_b \lambda_4 e^{-2\pi/\alpha} \varphi \varphi' + \dots
    \label{eq:Vsmallins}
\end{equation}
which give rise to a potential generated by small instantons 
\begin{equation}
	V(a) \simeq  y_t y_b e^{-\frac{2\pi}{\alpha}} v_9^2 \left( \frac{\lambda_4 v_\varphi v_{\varphi'}}{(4\pi)^2} + \frac{v_{\rm EW}^2 \sin 2\beta}{4} \right) \left(1 - \cos\left(2\frac{a}{f_a}\right)\right) + \Lambda_{\rm QCD}^4 \left(1 - \cos\left(6\frac{a}{f_a}\right)\right).
\label{eq:axionVquartic}
\end{equation}

We note that this left-right version of the $SU(9)$ color-flavor unification model is motivated also from the top down, as it can arise starting from the complete fermion unification model $SU(12) \times SU(2)_L \times SU(2)_R$ \cite{Allanach:2021bfe,Davighi:2022fer,Davighi:2022qgb,Delgado:2026a}.

\begin{figure}[th]
\centering
\hspace*{-5mm}
\includegraphics[width=0.69\textwidth]{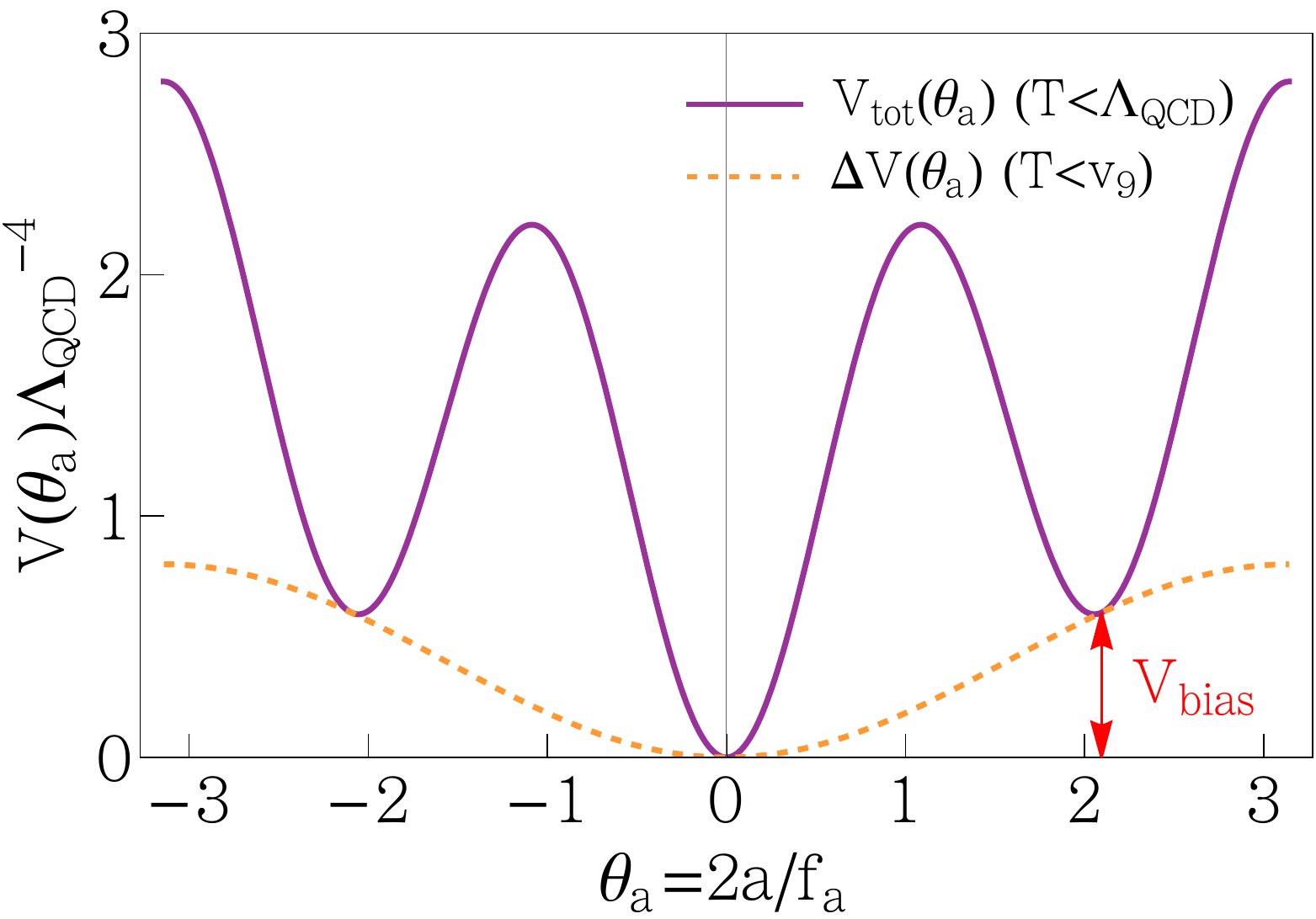}
\caption{Axion potential in the unit of $\Lambda_{\rm QCD}^{4}$ for $-\pi\leq\theta_{a}\equiv 2a/f_{a}\leq\pi$. The purple line shows the net axion potential contributed both by QCD and the constrained instanton of $SU(9)$ (orange dashed line). The induced bias is $V_{\rm bias}=V_{\rm tot}(\theta_{a}=\pm2\pi/3)$. 
}
\vspace*{-1.5mm}
\label{fig:Vtotbias}
\end{figure}

\section{Gravitational Waves from Domain Wall Collapse}\label{sec:GW}

In this section, we discuss the gravitational wave signature that could possibly come from the collapse of the domain walls. To this end, in Appendix~\ref{sec:AppendA} we briefly summarize how the peak frequency $f_{p,0}$ and strength of the gravitational wave spectrum at $f=f_{p,0}$ are obtained following \cite{Saikawa:2017hiv}. We focus on gravitational waves from domain wall annihilation since the contribution from cosmic strings is expected to be subdominant relative to that from domain wall annihilation. For the stochastic gravitational wave background produced by axion string, see, e.g. \cite{Gorghetto:2021fsn}.

For both cubic and quartic models, we can parametrize the height of the leading correction to the QCD axion potential by $\Lambda_{9}^{4}$, and then the net axion potential can be written as 
\begin{equation}
V_{\rm tot}(a)=\Lambda_{\rm QCD}^{4}\left[1-\cos\left(6\frac{a}{f_{a}}\right)\right]+\Delta V\left(a\right)\,,  
\label{eq:Vtot}
\end{equation}
where $\Delta V(a/f_{a})\equiv\Lambda_{9}^{4}(1-\cos(2a/f_{a}))$ is the contribution from the constrained instanton in $SU(9)/[SU(3)^2/\mathbb{Z}_3]$. In Fig.~\ref{fig:Vtotbias}, we show $V_{\rm tot}(a)$ for the much exaggerated example of $\Lambda_{9}^{4}/\Lambda_{\rm QCD}^{4}=0.4$ for illustration. 
The purple line shows $V_{\rm tot}(a)$ whereas the orange dashed line corresponds to $\Delta V(a)$. The degeneracy of the three vacua around the minimal string produced at PQ phase transition is protected by a $\mathbb{Z}_3$ non-invertible shift symmetry and broken by $\Delta V(a)$. From here on, we treat the ratio $\Lambda_{9}^{4}/\Lambda_{\rm QCD}^{4}$ as a free parameter, remaining agnostic about details of a $SU(9)$ breaking scale $v_{9}$ and the gauge coupling.

While the axion strings form at the PQ phase transition ($T\simeq f_{a}$), domain walls (DW) do not form until around the QCD phase transition ($T\simeq\Lambda_{\rm QCD}$). Although $\Delta V(a)\neq0$ is present before $T\simeq\Lambda_{\rm QCD}$ is reached, the axion mass remains smaller than $H$ until the QCD phase transition. 
Then the axion begins dynamically evolving due to $m_{a}\sim\Lambda_{\rm QCD}^2 /f_{a}\gtrsim H$, and domain wall formation takes place. Given the discussion in Sec.~\ref{sec:cubic} and \ref{sec:quartic}, in our scenario three DWs are expected to end on a minimal-winding axion string. The bias $V_{\rm bias}\equiv V_{\rm tot}(\pm 2\pi/3)=\Delta V(\pm 2\pi/3)$ introduced by $SU(9)$ constrained instanton causes the pressure force $p_{V}\simeq V_{\rm bias}$, and when it wins against the tension force $p_{T}\sim\mathcal{A}\sigma/t$ where $\sigma$ is the domain wall tension in Eq.~\ref{eq:tension} and $\mathcal{A}\simeq0.8$~\cite{Hiramatsu:2013qaa}, DWs are annihilated and disappear, radiating gravitational wave (GW) and the axion quanta.

\begin{figure}[t]
\centering
\hspace*{-5mm}
\includegraphics[width=0.69\textwidth]{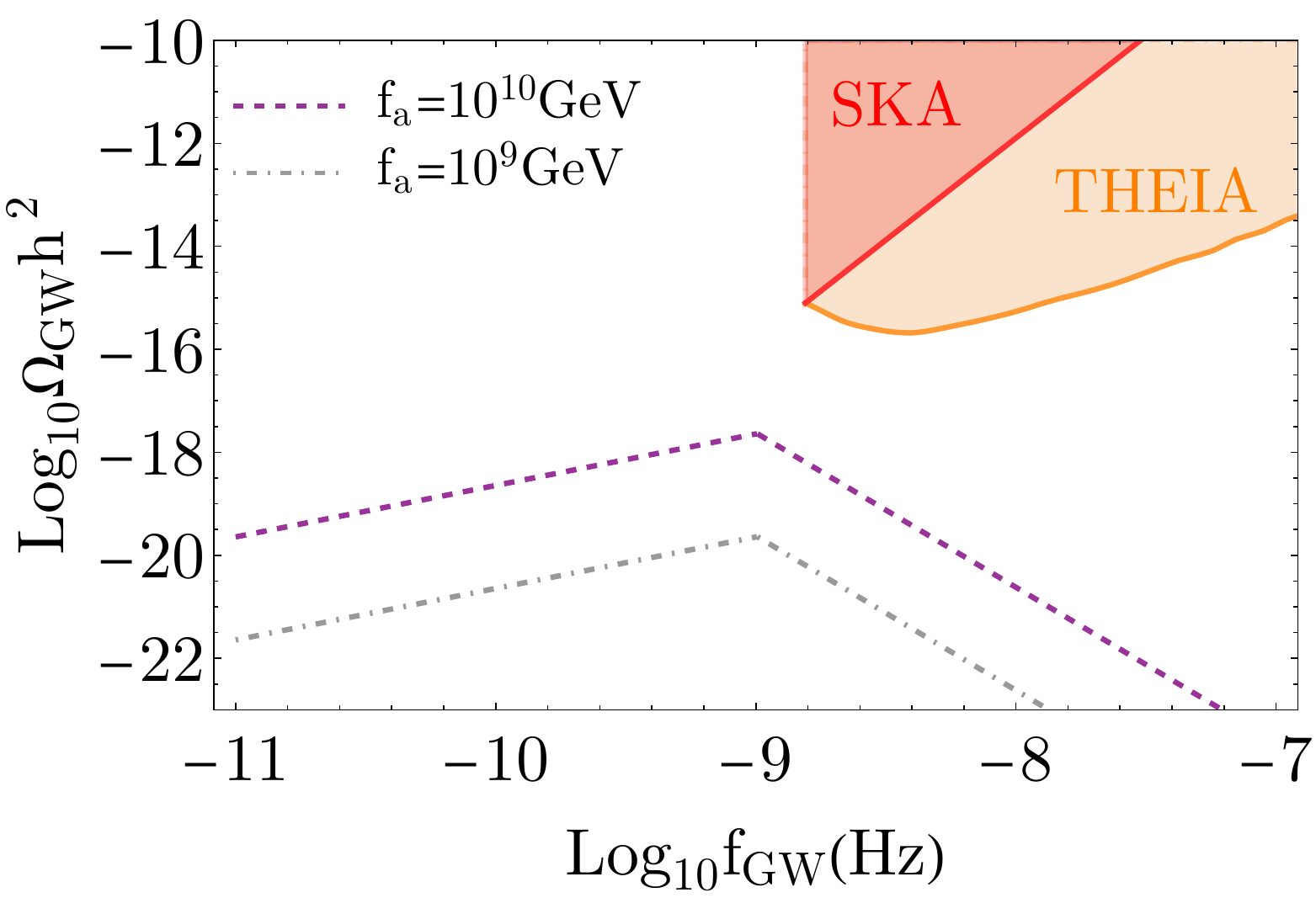}
\caption{Gravitational wave spectrum from the domain wall collapse for each specified $f_{a}$ and $T_{\rm ann}=10{\rm MeV}$. The purple dashed (gray dotdashed) line corresponds to $\Omega_{\rm GW,0}h^{2}$ for $f_{a}=10^{10}{\rm GeV}$ ($10^{9}{\rm GeV}$). The orange and red shaded regions are the projected sensitivities of the future detectors, SKA
~\cite{Carilli:2004nx,Janssen:2014dka,Weltman:2018zrl} and THEIA~\cite{thetheiacollaboration2017theiafaintobjectsmotion}.}
\vspace*{-1.5mm}
\label{fig:GWDW}
\end{figure}

From $p_{T}\simeq p_{V}$, we can estimate the temperature $T_{\rm ann}$ of the thermal bath when domain wall collapse occurs. Using the surface tension $\sigma=(8/N_{\rm DW}^{2}) m_{a}f_{a}^{2}$, 
we find
\begin{equation}
T_{\rm ann}\simeq3.41\times10^{-2}{\rm GeV}C_{d}^{-1/2}\mathcal{A}^{-1/2}\left(\frac{\Lambda_{\rm QCD}^{2}}{{\rm GeV}^{2}}\right)^{-1/2}\left(\frac{f_{a}}{10^{9}{\rm GeV}}\right)^{-1/2}\left(\frac{V_{\rm bias}}{{\rm MeV}^{4}}\right)^{1/2}\,.
\label{eq:Tann}
\end{equation}
where $C_{d}=\mathcal{O}(1)$ depends on $N_{\rm DW}$ and the definition of the decay time. We take $C_{d}=5$ by following $10\%$ criterion of the decay time in \cite{Kawasaki:2014sqa}. As the axions from the domain wall collapse contribute to the relic abundance of the axion dark matter, $f_{a}$ is constrained to be $f_{a}\lesssim\mathcal{O}(10^{10}){\rm GeV}$ to avoid overclosing the universe~\cite{Klaer:2017ond,Gorghetto:2018myk,Buschmann:2019icd}. On the other hand, the requirement that the domain wall annihilation be completed before BBN era leads to $T_{\rm ann}\gtrsim 10{\rm MeV}$. From these constraints on $f_{a}$ and $T_{\rm ann}$, we find the lower bound on $V_{\rm bias}$
\begin{equation}
V_{\rm bias}=\Delta V(\pm 2\pi/3)=\frac{3}{2}\Lambda_{9}^{4}\,\gtrsim\, (0.54{\rm MeV})^{4}C_{d}\mathcal{A}\left(\frac{\Lambda_{\rm QCD}^{2}}{{\rm GeV}^{2}}\right)\left(\frac{f_{a}}{10^{9}{\rm GeV}}\right)\,. 
\label{eq:Vbias}
\end{equation}
Therefore, for $f_{a}\gtrsim10^{9}{\rm GeV}$ that avoids to trigger too large a stellar cooling~\cite{Raffelt:2006cw,ParticleDataGroup:2024cfk} and $\Lambda_{\rm QCD}\simeq300{\rm MeV}$, $\Lambda_{9}^{4}\gtrsim0.02{\rm MeV}^{4}$ is required for the successful BBN. 

From $T_{\rm ann}\propto\sqrt{V_{\rm bias}/f_{a}}$ and the spectrum of the gravitational wave from the domain wall collapse $\Omega_{\rm GW,0}h^{2}\propto f_{a}^{2}/T_{\rm ann}^{4}$ in Eq.~\ref{eq:peakamp}, one can infer that the most optimistic GW signal consistent with $T_{\rm ann}\gtrsim10{\rm MeV}$ and $\Omega_{a}\lesssim\Omega_{\rm cdm}$ can be obtained for  $f_{a}\simeq10^{10}{\rm GeV}$. For $T_{\rm ann}=10{\rm MeV}$, we show $\Omega_{\rm GW,0}h^{2}$ for two choices of $f_{a}$ consistent with $\Omega_{a}\lesssim\Omega_{\rm cdm}$ and the stellar cooling bound on $f_{a}$ in Fig.~\ref{fig:GWDW}. As the peak frequency is sensitive to $T_{\rm ann}$ only, both spectra are peaked at $f_{p,0}\simeq10^{-9}{\rm Hz}$. We see that unfortunately $\Omega_{\rm GW,0}h^{2}$ is not large enough to be probed by the future detectors, SKA
~\cite{Carilli:2004nx,Janssen:2014dka,Weltman:2018zrl} and THEIA~\cite{thetheiacollaboration2017theiafaintobjectsmotion}, given our having to reconcile the scenario with successful BBN and dark matter relic abundance. This feature is common to models where QCD axion domain walls annihilate due to a bias potential. 

\section{Conclusions}

In this work we have demonstrated the phenomenological importance of the global structure of symmetry groups and the topology of field space in multiple ways---in the determination of the axion mode, in the spectrum of cosmic strings and domain walls, in the spectrum of fractional instantons, and in guiding model building to solve the domain wall problem. It is worth thinking through other popular BSM theories with this in mind and considering whether global structure has important lessons for these models as well.

There are many directions ripe for further investigation. How to identify the axion mode in a general theory with many scalars and many gauge factors seems an important question. 
It would be interesting to further study the right-handed version of the PQWW axion we embedded the quartic axion model in, and to write a fully realistic model thereof. And of course there is much more interesting phenomenology to uncover in quark color-flavor unification, both at terrestrial experiments and in the early universe.
Finally, the type of global structure we have considered here is not the most general case, and it would be useful to understand the full interplay with the Lazarides-Shafi mechanism.

\section*{Acknowledgements}

We are grateful to Antonio Delgado, Adam Martin, and  Lian-Tao Wang for helpful conversations. 
The work of GC is supported in part by DOE grant DE-SC0011842 at the University of
Minnesota.
The work of SH is supported by the National Research Foundation of Korea (NRF) Grants RS-2023-00211732, RS-2024-00405629, RS-2026-25481264, and by the Samsung Science and Technology Foundation under Project Number SSTF-BA2302-05, and by the POSCO Science Fellowship of POSCO TJ Park Foundation. 
The work of SK is partially supported by the National Science Foundation under grant PHY-2412701.
SH and SK are grateful for hospitality during the CERN-CKC TH Institute 2024, the KITP program `What is Particle Theory?' (supported by the National Science Foundation under Grant No. NSF PHY-1748958), the KAIST workshop `Symmetries in Quantum Field Theory and Particle Physics', and the IHES workshop `Symmetry and Topology in Particle Physics'. 

\appendix

\section{Gravitational Wave from Domain Walls}
\label{sec:AppendA}
When $N_{\rm DW}\geq1$, the string-wall system is stable provided $\mathbb{Z}_{N_{\rm DW}}$ is an exact symmetry. When $m_{a}\simeq H$ is satisfied, the axion starts to oscillate and domain walls are formed with the spontaneous breaking of $\mathbb{Z}_{N_{\rm DW}}$. In this case, $N_{\rm DW}$ domain walls end on a single axion string. If particles residing in the SM thermal bath interact with the axions, there is a thermal friction force $p_{F}\simeq\Delta pn\sim vT^{4}$ on the domain walls, where $\Delta p\sim vT$ is a momentum transfer and $n\sim T^{3}$ is a number density of the particles in the thermal bath. This force is balanced by the tension force $p_{T}\sim\sigma/R_{\rm wall}$ where $R_{\rm wall}$ is the curvature radius of a domain wall and $\sigma$ is the tension of the domain wall separating degenerate minima which is given by 
\begin{equation}
\sigma=\frac{8}{N_{\rm DW}^{2}}m_{a}f_{a}^{2}\,.
\label{eq:tension}
\end{equation}

After the friction force becomes inefficient, the string-domain wall system enters the scaling regime where the wall energy density gets diluted as $\rho_{w}(t)\propto t^{-1}$. When the universe is in the radiation-dominated era, this implies $\rho_{w}(t)\propto a^{-2}$, and thus the dilution of $\rho_{w}$ is slower than that of matter ($\rho_{m}\propto a^{-3}$) and radiation ($\rho_{\rm rad}\propto a^{-4}$). Therefore, in case where domain walls remain stable, they quickly dominate the energy of the universe, which gives rise to the uwanted rapid expansion of the universe. In our set-up, domain walls become unstable as the degenerate vacua are lifted up thanks to the breaking of the non-invertible $\mathbb{Z}_{3}$ symmetry that is caused by the additional contribution to the axion potential from the small instantons of the UV group.

The potential bias $V_{\rm bias}$ induces a volume pressure force $p_{V}\sim V_{\rm bias}$ acting on the wall and domain walls get annihilated once $p_{T}\simeq p_{V}$. Since the frequency of the GWs from the wall collapse is determined by the horizon size, i.e. $f_{p}(t_{\rm ann})=H(t_{\rm ann})$, after redshifting $f_{p}(t_{\rm ann})$ to today, we obtain the following peak frequency today of the GWs from the domain wall collapse 
\begin{equation}
    f_{p,0}=\frac{H(t_{\rm ann})a_{\rm ann}}{a_{0}}=10^{-9}{\rm Hz}\left(\frac{g_{*}(T_{\rm ann})}{10}\right)^{1/2}\left(\frac{g_{*s}(T_{\rm ann})}{10}\right)^{-1/3}\left(\frac{T_{\rm ann}}{10^{-2}{\rm GeV}}\right)\,,
\end{equation}
where $T_{\rm ann}$ can be read from Eq.~\ref{eq:Tann} and $g_{*s}(T_{\rm ann})$ is the effective number of the relativistic degrees of freedom for the entropy density at $T=T_{\rm ann}$.

GW spectrum is defined as 
\begin{equation}
    \Omega_{\rm GW}(t,f)=\frac{1}{\rho_{c}(t)}\frac{d\rho_{\rm GW}(t)}{d\ln f}\,,
\end{equation}
where $\rho_{c}(t)$ and $\rho_{\rm GW}$ are the critical energy density of the universe and the energy density of GW at a time $t$, respectively. At the time of annihilation, the spectrum at the peak is given by~\cite{Hiramatsu:2013qaa}
\begin{equation}
 \Omega_{\rm GW}(t_{\rm ann},f_{p}(a_{\rm ann}))=\frac{1}{\rho_{c}(t_{\rm ann})}\left(\left.\frac{d\rho_{\rm GW}(t_{\rm ann})}{d\ln f}\right|_{f=f_{p}(a_{\rm ann})}\right)=\frac{\tilde{\epsilon}_{\rm GW}G\mathcal{A}^{2}\sigma^{2}}{\rho_{c}(t_{\rm ann})}\,,
\end{equation}
where $\tilde{\epsilon}_{\rm GW}=d\epsilon_{\rm GW}/d\ln k$ with $\epsilon_{\rm GW}$ a parameter quantifying the GW emission efficiency and $G=(8\pi M_{P}^{2})^{-1}$ is a Newton's constant with $M_{P}=2.4\times10^{18}{\rm GeV}$ the reduced Planck mass. During the scaling regime, $\tilde{\epsilon}_{\rm GW}$ was numerically shown to remain almost constant and estimated to be $\tilde{\epsilon}_{\rm GW}\simeq0.7\pm0.4$. Then from 
\begin{equation}
    \Omega_{\rm GW,0}=\frac{1}{\rho_{c,0}}\frac{d\rho_{\rm GW,0}}{d\ln f}=\frac{\rho_{c}(t_{\rm ann})}{\rho_{c,0}}\left(\frac{a(t_{\rm ann})}{a_{0}}\right)^{4} \frac{1}{\rho_{c}(t_{\rm ann})}\left(\frac{d\rho_{\rm GW}(t_{\rm ann})}{d\ln f}\right)\,,
\label{eq:OmegaGW0}
\end{equation}
where the second equality follows from $\rho_{\rm GW}\propto a^{-4}$, using the entropy conservation from $t_{\rm ann}$ to $t_{0}$, $\rho_{c}(t_{\rm ann})/\rho_{c,0}=(g_{*}(T_{\rm ann})/g_{*,0})(T_{\rm ann}^{4}/T_{0})^{4}$ and Eq.~(\ref{eq:OmegaGW0}), one obtains at $f=f_{p,0}$~\cite{Saikawa:2017hiv}
\begin{equation}
\Omega_{\rm GW,0}h^{2}=7\times10^{-18}\tilde{\epsilon}_{\rm GW}\mathcal{A}^{2}\left(\frac{g_{*s}(T_{\rm ann})}{10}\right)^{-4/3}\left(\frac{\Lambda_{\rm QCD}^{2}}{{\rm GeV}^{2}}\right)^{2}\left(\frac{f_{a}}{10^{9}{\rm GeV}}\right)^{2}\left(\frac{T_{\rm ann}}{10^{-2}{\rm GeV}}\right)^{-4}\,,
\label{eq:peakamp}
\end{equation}
where we defined $\Omega_{\rm GW,0}\equiv \Omega_{\rm GW}(t_{0})$ and $h\simeq0.7$ is defined via the current Hubble expansion rate $H_{0}=100h{\rm km}/{\rm sec}/{\rm Mpc}$. Given the peak amplitude in Eq.~\ref{eq:peakamp}, the frequency dependence suggested by the numerical simulations reads $\Omega_{\rm GW,0}\propto f^{3}$ for $f<f_{p}$ and $\Omega_{\rm GW,0}\propto f^{-1}$ for $f>f_{p}$~\cite{Hiramatsu:2013qaa}.

\bibliographystyle{jhep}
\bibliography{nipqdfsz}

\end{document}